\documentclass{ptephy_v1}

\preprintnumber{XXXX-XXXX} 

\usepackage{bm} 
\usepackage{aas_macros} 

\input{colordvi.tex} 

\begin{document}

\title{Delta-map method to remove CMB foregrounds with spatially varying
spectra}


\author[1,2]{Kiyotomo Ichiki}
\affil[1]{Graduate School of Science, Division of Particle and
Astrophysical Science, Nagoya University, Chikusa-ku, Nagoya, 464-8602, Japan}
\affil[2]{Kobayashi-Maskawa Institute for the Origin of Particles and the
Universe, Nagoya University, Chikusa-ku, Nagoya, 464-8602, Japan
\email{ichiki@a.phys.nagoya-u.ac.jp}}

\author[3]{Hiroaki Kanai}
\affil[3]{Graduate School of Engineering, Yokohama National University,
79-5 Tokiwadai, Hodogaya-ku, Yokohama 240-8501 JAPAN}

\author[4]{Nobuhiko Katayama}
\affil[4]{Kavli Institute for the Physics and Mathematics of the Universe
(Kavli IPMU, WPI), Todai Institutes for Advanced Study, The University
of Tokyo, Kashiwa 277-8583, Japan}

\author[4,5]{Eiichiro Komatsu}
\affil[5]{Max Planck Institute for Astrophysics, Karl-Schwarzschild-Str. 1,
	D-85748 Garching, Germany}


\begin{abstract}%
 We extend the internal template foreground removal method
 by accounting for spatially varying spectral parameters such as the
 spectral indices of synchrotron and dust emission and the dust
 temperature. As the previous algorithm had to assume that the spectral
 parameters are uniform over the full sky (or some significant fraction
 of the sky), it resulted in a bias in the tensor-to-scalar ratio
 parameter $r$ estimated from foreground-cleaned polarization maps of the cosmic
 microwave background (CMB). The new algorithm, ``Delta-map method'',
 accounts for spatially varying spectra to first order in
 perturbation. The only free parameters are the cosmological parameters
 such as $r$ and the sky-averaged foreground parameters.
 We show that a cleaned CMB map is the maximum likelihood
 solution to first order in perturbation, and derive the posterior
 distribution of $r$ and the sky-averaged foreground parameters using Bayesian
 statistics. Applying this to realistic
 simulated sky maps including polarized CMB, synchrotron and thermal
 dust emission, we find that the new algorithm removes the bias in $r$
 down to undetected level in our noiseless simulation ($r\lesssim
 4\times 10^{-4}$). We show that the frequency decorrelation of
 polarized foregrounds due to averaging of spatially varying spectra can
 be accounted for to first order in perturbation by using slightly
 different spectral parameters for the Stokes $Q$ and $U$ maps. Finally,
 we show that the effect of polarized anomalous microwave emission on
 foreground removal can be absorbed into the curvature parameter of the
 synchrotron spectrum.
\end{abstract}
\subjectindex{xxxx, xxx}
\maketitle
\section{Introduction}
Search for the signature of primordial gravitational waves from the
early Universe in the B-mode polarization of the cosmic microwave
background (CMB) \cite{2016ARAA..54..227K} is limited already by
accuracy of the foreground removal procedure
\cite{2015PhRvL.114j1301B}. Among various methods for removing the
foreground emission (see
\cite{2008A&A...491..597L,2018JCAP...04..023R,2018arXiv180706208P} and
references therein), we focus on the so-called ``internal template''
method
\cite{2007ApJS..170..335P,2009MNRAS.397.1355E,2011ApJ...737...78K,2012MNRAS.420.2162F,2016MNRAS.459..441F}. The
basic idea behind this method is simple. Consider that we have maps at
three frequencies: low, middle, and high. Say 40, 100, and 240~GHz. We
know that the low and high frequency data are dominated by synchrotron
and thermal dust emission, respectively \cite{2014PTEP.2014fB109I};
thus, the simplest way to remove them is to fit the low- and
high-frequency maps to and subtract from the map at the middle
frequency. A cleaned map may be written as
$x=([Q,U]_{100}+\alpha_{40}[Q,U]_{40}+\alpha_{240}[Q,U]_{240})/(1+\alpha_{40}+\alpha_{240})$,
where $\alpha_{40}$ and $\alpha_{240}$ are fitting coefficients and $Q$
and $U$ are the Stokes parameters in thermodynamic temperature units.
Fitting
can be done in pixel, Fourier (spherical harmonics), or wavelet
space. This method is called a {\it template} because the low and high
frequency data are used as templates of synchrotron and thermal dust
emission, respectively, and is called {\it internal} because no external
data sets are used.

All of the previous implementations of this method assumed that the
fitting coefficients are uniform over some fitted regions. In some study
the fitted region was taken to be the whole sky
\cite{2007ApJS..170..335P,2009MNRAS.397.1355E,2012MNRAS.420.2162F,2016MNRAS.459..441F},
whereas in the other some large pixels \cite{2011ApJ...737...78K}. In this paper, we build
on our previous work \cite{2011ApJ...737...78K}, in
which we find that the internal template method yields a small but
non-negligible bias in the tensor-to-scalar ratio parameter $r$ of order
$10^{-3}$, and that the origin of the bias is the spatial variation of
the fitting coefficients, i.e., spatial variation of the spectral
parameters of the foreground emission.

Motivated by this finding, we extend the
internal template method by accounting for spatial variations of the
spectra. If we have two high (or low) frequency channels as dust (or
synchrotron) templates instead of one, we should be able to solve for
both the mean and variation in one spectral parameter, e.g., the dust
(or synchrotron) spectral index. We show that this is indeed the case.
In general, we need $N+1$ channels to account for spatial variations of
$N$ spectral parameters of one foreground component.

Armed with the new method, which we shall call the ``Delta-map method''
throughout this paper, we investigate two hotly debated topics of
the polarized foreground: treatments of possible polarization in
the anomalous microwave emission (AME) and the frequency decorrelation
due to averaging of spatially varying foreground spectra.

This paper is organized as follows.
We introduce the Delta-map method in Sect.~\ref{sec:deltamap}, and
explain how to implement it in Sect.~\ref{sec:implementation}.
In Sect.~\ref{sec:bayesian}, we interpret the Delta-map method within
the Bayesian framework. In Sect.~\ref{sec:validation}, we apply
this to realistic foreground maps for validation. In
Sect.~\ref{sec:application}, we show how to treat AME and frequency
decorrelation. In Sect.~\ref{sec:noise}, we study how instrumental
noise affects the results. We conclude in Sect.~\ref{sec:conclusions}.

\section{Delta-map method}
\label{sec:deltamap}
\subsection{General formula}
\label{sec:2.1}
Without loss of generality, let us consider one foreground component
with $N$ spectral parameters, and model the observed Stokes parameters $Q$ and
$U$ in thermodynamic temperature units at a given frequency $\nu$ as
\begin{equation}
 [Q,U]_\nu(\hat{n}) =   \mbox{CMB}(\hat{n})
  +g_\nu D_{\nu}(p^I(\hat{n}))[Q^f,U^f]_{\nu_\ast}(\hat{n})
  +N_{\nu}(\hat{n})~,
\label{eq:QUmodel}
\end{equation}
where $\mbox{CMB}(\hat{n})$ and $N_\nu(\hat{n})$ denote the CMB signal and
instrumental noise towards a sky direction $\hat{n}$ in thermodynamic
units, respectively, and 
$[Q^f,U^f]_{\nu_\ast}(\hat{n})$ denotes the foreground component at some
reference frequency $\nu_\ast$ in brightness temperature units. As the
dependence on $\nu_*$ cancels in the final result, $[Q^f,U^f]_{\nu_\ast}$
appears only as an auxiliary variable. For simplicity we have not
written explicitly experiment-specific effects such as a gain calibration
factor, beam smearing, pixel window function, or any other
smoothing/filtering that may be applied to data. Such operations will be common
to all sky signals at the same frequency channel.

The coefficient $g_\nu$ is given by
\begin{equation}
g_\nu \equiv \frac{(e^x-1)^2}{e^xx^2} ~~ \mbox{with} ~~
 x\equiv \frac{h\nu}{k_BT_{\rm CMB}}~,
 \label{eq:Fnu}
\end{equation}
which converts the units from brightness temperature to the CMB thermodynamic
temperature ($T_{\rm CMB}=2.725$~K). While this expression does not take
into account averaging over the bandpass of detectors, we shall include
it in Sect.~\ref{sec:bandpass}.

The function $D_\nu$ denotes the frequency dependence of the foreground
emission in brightness temperature units, and $p^I(\hat{n})$ a set of parameters
($I=1,2,\cdots, N$) that characterize the emission law and can vary
across the sky. For example, if we consider a power-law
foreground emission component, then $N=1$ and $p^1(\hat{n})$ is a
power-law spectral index $\beta(\hat{n})$. Thus, $D_\nu =
\left({\nu}/{\nu_\ast}\right)^{\beta(\hat{n})}$.
Decomposing the spectral parameters to the mean value and perturbation
as $p^I(\hat{n}) = \bar{p}^I+\delta p^I(\hat{n})$, we obtain
\begin{equation}
D_\nu(p^I(\hat{n}))[Q^f,U^f]_{\nu_\ast}(\hat{n}) 
 = D_\nu(\bar{p}^I)[Q^f,U^f]_{\nu_\ast}(\hat{n})
 + \sum_{J=1}^N D_{\nu,J}\delta
 p^J(\hat{n})[Q^f,U^f]_{\nu_\ast}(\hat{n})+{\cal O}(\delta p^2)\,,
\label{eq:QUmodel2}
\end{equation}
where we wrote $D_{\nu,J}\equiv \left({\partial D_\nu}/{\partial
		 p^J}\right)_{p^J = \bar{p}^J}$.

The central assumption we make throughout this paper is that we truncate
the expansion at first order in $\delta p^I$ \cite{1999NewA....4..443B,2016ApJ...829..113S}. Therefore, when a
foreground component is parameterized by $N$ parameters that vary across
the sky only weakly, we can interpret the emission as a superposition of $(N+1)$ template maps; namely
		 $[Q^f,U^f]_{\nu_\ast}(\hat{n})$ and $\delta
		 p^I(\hat{n})[Q^f,U^f]_{\nu_\ast}(\hat{n})$
		 $(I=1,2,\cdots,N)$. 
If the ${\cal O}(\delta p^2)$ term is significant compared with the first two
terms, it may cause a bias in $r$ as foreground residuals. The bias becomes
greater when the foreground channels are taken further away from the CMB
channel. We find, however, that the bias is negligible for the band
specification and realistic simulated sky maps used in this work, as we shall
show later.
		 
		 To subtract the foreground component in Eq.~(\ref{eq:QUmodel})
with help of Eq.~(\ref{eq:QUmodel2}), we form a linear combination of
$(N+2)$ maps at different frequencies as 
\begin{eqnarray}
[Q,U]_{\nu_{\rm CMB}}(\hat{n}) + \sum_{j=1}^{N+1}\alpha_j[Q,U]_{\nu_j}(\hat{n})
 &=&
\left(1+\sum_{j=1}^{N+1}\alpha_j \right)\mbox{CMB}(\hat{n}) \nonumber \\
&+&
\left(g_{\nu_{\rm CMB}}D_{\nu_{\rm CMB}}(\bar{p}^I)+\sum_{j=1}^{N+1} \alpha_j g_{\nu_j}D_{\nu_j}(\bar{p}^I)  \right)[Q^f,U^f]_{\nu_\ast}(\hat{n}) \nonumber \\
&+&
\sum_{J=1}^{N}\left(
g_{\nu_{\rm CMB}} D_{\nu_{\rm CMB},J}
+ \sum_{j=1}^{N+1} \alpha_j g_{\nu_j}D_{\nu_j,J} 
\right) \delta p^J (\hat{n})[Q^f,U^f]_{\nu_\ast}(\hat{n})\nonumber \\
 &+&
N_{\nu_{\rm CMB}}(\hat{n})+\sum_{j=1}^{N+1} \alpha_j N_{\nu_j}(\hat{n})~,
\end{eqnarray}
where $\alpha_j$ are fitting coefficients, $\nu_{\rm CMB}$ denotes a frequency
employed as a CMB channel, and $\nu_j$ the foreground channels.
The fitting coefficients are found by setting the foreground terms in
the above equation (i.e., the second and third lines) to zero:
\begin{eqnarray}
 g_{\nu_{\rm CMB}}D_{\nu_{\rm CMB}}(\bar{p}^I)+\sum_{j=1}^{N+1} g_{\nu_{j}}D_{\nu_j}(\bar{p}^I) \alpha_j &=& 0\,, \label{eq:alphas1}\\
g_{\nu_{\rm CMB}}D_{\nu_{\rm CMB},J}
+ \sum_{j=1}^{N+1} g_{\nu_{j}}D_{{\nu_j},J}\alpha_j  &=& 0\,,
\quad \mbox{for} ~ J=(1,2,\cdots,N).
\label{eq:alphas2}
\end{eqnarray}
These equations can be cast into a matrix form as ${\bm
A}\vec{\alpha}=-\vec{d}_{\nu_{\rm CMB}}$, where
\begin{eqnarray}
\vec{\alpha} &\equiv&(\alpha_1, \alpha_2, \cdots, \alpha_{N+1})^{\rm T}
 ~,\\
 \label{eq:dfunc}
\vec{d}_{\nu} &\equiv& g_{\nu}\left(D_{\nu}(\bar{p}^I),
			       D_{\nu,1},\cdots,
			       D_{\nu,N} \right)^{\rm T}~,\\
{\bm A}&\equiv& (\vec{d}_{\nu_{1}},\vec{d}_{\nu_{2}},\cdots,\vec{d}_{\nu_{N+1}})~.
\end{eqnarray}
The coefficients $\alpha_j$ can be obtained numerically by $-{\bm
A}^{-1}\vec{d}_{\nu_{\rm CMB}}$\footnote{The explicit form of ${\bm A}$ is
\begin{equation}
 {\bm A} =
\begin{pmatrix}
g_{\nu_1} D_{\nu_1} & g_{\nu_2}D_{\nu_2} & \cdots & g_{\nu_{N+1}}D_{\nu_{N+1}}\\
 g_{\nu_1}D_{\nu_1,1} &
 g_{\nu_2}D_{\nu_2,1} & \cdots &
 g_{\nu_{N+1}} D_{\nu_{N+1},1}\\
 g_{\nu_1} D_{\nu_1,2} &
 g_{\nu_2} D_{\nu_2,2} &
 \cdots &  g_{\nu_{N+1}} D_{\nu_{N+1},2}\\
 \vdots & \vdots & \ddots & \vdots\\
 g_{\nu_1} D_{\nu_1,N} &
 g_{\nu_2} D_{\nu_2,N} & \cdots &
 g_{\nu_{N+1}} D_{\nu_{N+1},N}
\end{pmatrix}
\end{equation}}.
Once the coefficients are determined, we obtain a foreground-cleaned CMB
map $\vec{x}$ as
\begin{equation}
 \vec{x} = \frac{[Q,U]_{\nu_{\rm CMB}}(\hat{n}) +
  \sum_{j=1}^{N+1}\alpha_j[Q,U]_{\nu_j}(\hat{n})}{1+\sum_{j=1}^{N+1}\alpha_j}\,.
\label{eq:cleanedmap}
\end{equation}
Since this is a linear combination of maps, we can construct a cleaned
CMB in pixel, spherical harmonics, or wavelet space.

Multi-component foregrounds can be incorporated straightforwardly by
expanding the entries of ${\bm{A}}$. For example, if we
use $D_\nu$ for dust and $S_\nu$ for synchrotron, we may define a new
vector for synchrotron as
\begin{equation}
 \vec{s}_{\nu} \equiv g_{\nu}\left(S_{\nu}(\bar{p}^J),
			       S_{\nu,1},\cdots,
			       S_{\nu,M} \right)^{\rm T}~,
\end{equation}
where $M$ is the number of parameters characterizing synchrotron,
and write the matrix $\bm{A}$ as 
\begin{equation}
 {\bm A}=
  (\vec{d}_{\nu_{1}},\vec{d}_{\nu_{2}},\cdots,\vec{d}_{\nu_{N+1}},
 \vec{s}_{\nu_{1}},\vec{s}_{\nu_{2}},\cdots,\vec{s}_{\nu_{M+1}}
)~.
\end{equation}
There are $N+M+2$ coefficients to be determined, and are given by 
$\vec{\alpha}=-\bm{A}^{-1}(\vec{d}_{\nu_{\rm CMB}},\vec{s}_{\nu_{\rm
CMB}})^{\rm T}$. Thus, obtaining one cleaned CMB map requires at least
$N+M+3$ frequencies. In Sect.~\ref{sec:bayesian}, we derive the formula
for a cleaned CMB map when we have more frequency channels than this
minimum number. 

\subsection{Power law}
\label{sec:2.2.1}
To obtain a better feel for the new algorithm, let us work out three
simpler, concrete examples. 

First, we consider a single power-law spectrum with a spatially varying
spectral index:
\begin{equation}
 D_\nu(p^I(\hat{n}))
= D_\nu(\beta(\hat{n})) = \left(\frac{\nu}{\nu_\ast}\right)^{\beta(\hat{n})}~.
\end{equation}
The Stokes parameters become \cite{1999NewA....4..443B,2016ApJ...829..113S}
\begin{equation}
  [Q,U]_\nu(\hat{n})= \mbox{CMB}(\hat{n})+N_{\nu}(\hat{n})
   + g_{\nu}\left(\frac{\nu}{\nu_\ast}\right)^{\bar\beta}
  \left(1+\ln\left(\frac{\nu}{\nu_\ast}\right)\delta\beta(\hat{n}) \right)[Q^f,U^f]_{\nu_\ast}(\hat{n})
  +{\cal O}(\delta p^2)~,
\label{eq:simple_delta}
\end{equation}
where $\bar{\beta}$ is the mean spectral index. Eq.~(\ref{eq:dfunc}) becomes
\begin{eqnarray}
 \vec{d}_\nu = g_\nu
		\left(\frac{\nu}{\nu_\ast}\right)^{\bar{\beta}}\left(1
, \ln\left(\frac{\nu}{\nu_\ast}\right)\right)^{\rm T}~.
\end{eqnarray}
Therefore, the sum in Eq.~(\ref{eq:cleanedmap}) evaluates to
\begin{align}
\nonumber
 \sum_{j=1}^2\alpha_j\left[Q,U\right]_{\nu_j}(\hat n)
  =
  \frac{g_{\nu_{\rm CMB}}}{\ln(\nu_2/\nu_1)}
  &\left\{
   \frac1{g_{\nu_2}}\left(\frac{\nu_{\rm CMB}}{\nu_2}\right)^{\bar\beta}
   \ln(\nu_1/\nu_{\rm CMB})
   [Q,U]_{\nu_2}(\hat n)\right.\\
  & -\left.
   \frac1{g_{\nu_1}}\left(\frac{\nu_{\rm CMB}}{\nu_1}\right)^{\bar\beta}
   \ln(\nu_2/\nu_{\rm CMB})
   [Q,U]_{\nu_1}(\hat n)
  \right\}\,.
\label{eq:deltamapresultpowerlaw}
\end{align}
The reference frequency $\nu_*$ cancels in the final result, which is
required from the physical ground: a cleaned CMB map result should be
invariant under an arbitrary change in the reference frequency.
Adding this to $[Q,U]_{\nu_{\rm CMB}}(\hat n)$ removes the foreground
emission including spatial variation of the spectral index to first order in
$\delta\beta(\hat n)$. Note that we do not fit for $\delta\beta$ because
they are solved and eliminated. The only fitting parameter is the mean
spectral index $\bar\beta$.

As the above expression uses a difference between two weighted maps at
different frequencies, we started to call this map a ``Delta-map'',
which became the name of our algorithm. However, as we show in
Sect.~\ref{sec:bayesian},  our algorithm is much more general than
Eq.~(\ref{eq:deltamapresultpowerlaw}). A more appropriate (and general)
reason for the name ``Delta-map'' would be the fact that we expand and
truncate the spectral dependence of foregrounds at first order in
perturbation. Henceforth we shall adopt this reasoning for the name of
the algorithm.

\subsection{Running spectral index}
\label{sec:synchcurv}
Second, we allow the spectral index to vary slowly with frequency:
\begin{equation}
 D_\nu(p^I(\hat{n}))
= D_\nu(\beta(\hat{n}),C(\hat{n})) =
\left(\frac{\nu}{\nu_*}\right)^{\beta(\hat{n})+C(\hat{n})\ln\left({\nu}/{\nu_{\rm s}}\right)}~,
\end{equation}
where ${C}$ is the so-called curvature parameter or running of the
spectral index, $\bar C$ is its mean value, and $\nu_{\rm s}$ is a pivot
frequency of running index. There is no physical meaning for $\nu_{\rm
s}$, but it can be chosen such that it minimizes a correlation between
$\beta$ and $C$. 

The Stokes parameters become
\begin{eqnarray}
  [Q,U]_\nu(\hat{n}) &=& \mbox{CMB}(\hat{n})+N_{\nu}(\hat{n})\nonumber \\
\nonumber
  &+& g_{\nu}\left(\frac{\nu}{\nu_*}\right)^{\bar\beta+\bar
  C\ln\left({\nu}/{\nu_{\rm s}}\right)}
   \left[1+\ln\left(\frac{\nu}{\nu_*}\right)\left(\delta\beta(\hat{n})
    +\ln\left(\frac{\nu}{\nu_{\rm s}}\right)\delta C(\hat{n})\right)\right][Q^f,U^f]_{\nu_*}(\hat{n})\\
&+&{\cal O}(\delta p^2)\,,
\label{eq:runningQU}
\end{eqnarray}
 and
\begin{equation}
{\vec{d}}_\nu =  g_{\nu}\left(\frac{\nu}{\nu_*}\right)^{\bar\beta+\bar
 C\ln\left({\nu}/{\nu_{\rm
      s}}\right)}\left(1,\ln\left(\frac{\nu}{\nu_*}\right),\ln\left(\frac{\nu}{\nu_*}\right)\ln\left(\frac{\nu}{\nu_{\rm
		  s}}\right)\right)^{\rm T}
 \end{equation}
In this case, we need maps at three different frequencies to remove the
effects of spatially varying foreground parameters to first order in
$\delta\beta(\hat n)$ and $\delta C(\hat n)$.

\subsection{Modified Blackbody}
\label{sec:MBB}
Finally, we consider a modified blackbody (MBB) spectrum for dust emission:
\begin{equation}
D_\nu(p^I(\hat{n})) = D_\nu(\beta_{\rm d}(\hat{n}),T_{\rm d}(\hat{n})) = 
 \left(\frac{\nu}{\nu_*}\right)^{\beta_{\rm d}(\hat{n})+1}
 \frac{e^{x_{{\rm d}*}(\hat{n})}-1}{e^{x_{\rm d}(\hat{n})}-1}~,
\label{eq:MBBequation}
\end{equation}
where $x_{\rm d}(\hat{n})\equiv h\nu/k_BT_{\rm d}(\hat{n})$,
$x_{{\rm d}*}(\hat{n})\equiv h\nu_*/k_BT_{\rm d}(\hat{n})$, and
$T_{\rm d}(\hat{n})$ is the dust temperature. In the low-frequency limit,
$D_\nu \to (\nu/\nu_*)^{\beta_{\rm d}}$.
The Stokes parameters become
\begin{eqnarray}
 [Q,U]_\nu(\hat{n}) &=& \mbox{CMB}(\hat{n})+N_\nu(\hat{n}) 
+
 g_\nu\left(\frac{\nu}{\nu_*}\right)^{\bar{\beta}_{\rm
 d}+1}\frac{e^{\bar{x}_{{\rm d}*}-1}}{e^{\bar{x}_{\rm d}}-1}\nonumber \\
\nonumber
 & & \times
\left[
 1+\ln\left(\frac{\nu}{\nu_*}\right)\delta\beta_{\rm d}(\hat{n})
 +\left(\frac{\bar{x}_{\rm d}e^{\bar{x}_{\rm d}}}{e^{\bar{x}_{\rm d}}-1}
-\frac{\bar{x}_{{\rm d}*}e^{\bar{x}_{{\rm d}*}}}{e^{\bar{x}_{{\rm d}*}}-1}
  \right)\frac{\delta T_{\rm
 d}(\hat{n})}{\bar{T_{\rm d}}}
\right][Q^f,U^f]_{\nu_*}(\hat{n})\\
&+&{\cal O}(\delta p^2)\,,
\end{eqnarray}
where $\bar{T}_{\rm d}$ is the mean temperature,
$\bar{x}_{\rm d}\equiv h\nu/k\bar{T}_{\rm d}$, and 
\begin{equation}
 {\vec{d}}_\nu =
  g_\nu\left(\frac{\nu}{\nu_*}\right)^{\bar{\beta}_{\rm
  d}+1}\frac{e^{\bar{x}_{{\rm d}*}}-1}{e^{\bar{x}_{\rm d}}-1}
\left(1,\ln\left(\frac{\nu}{\nu_*}\right),
 \left(\frac{\bar{x}_{\rm d}e^{\bar{x}_{\rm d}}}{e^{\bar{x}_{\rm
  d}}-1}-\frac{\bar{x}_{{\rm d}*}e^{\bar{x}_{{\rm
  d}*}}}{e^{\bar{x}_{{\rm d}*}}-1}\right)\frac1{\bar{T}_{\rm d}}
\right)^{\rm T}~.
\end{equation}
In this case, we need maps at three different frequencies to remove the
effects of spatially varying foreground parameters to first order in
$\delta\beta_{\rm d}(\hat n)$ and $\delta T_{\rm d}(\hat n)$.

\subsection{Bandpass average}
\label{sec:bandpass}
The above formalism applies when we have detectors with a delta-function
response to a single frequency. In reality the sky signal must be
averaged within a given bandpass of detectors.

To incorporate the bandpass average, we first average an observed
intensity as
\begin{equation}
 \left<I(\nu)\right> \equiv \frac{\int
  W(\nu^\prime)I(\nu^\prime)d\nu^\prime}{\int W(\nu^\prime)d\nu^\prime}=
  \frac{1}{\Delta\nu}\int_{\nu-\Delta\nu/2}^{\nu+\Delta\nu/2}I(\nu^\prime)d\nu^\prime~,
\end{equation}
where the last equality holds when the bandpass function $W(\nu)$ is a
top hat function with a width of $\Delta \nu$. We then relate
$\left<I(\nu)\right>$ to the thermodynamic CMB temperature fluctuation
$\delta T$ as
\begin{equation}
 \left<I(\nu)\right> = \left<\frac{\partial B_\nu(T)}{\partial
  T}\right>\delta T~,
\end{equation}
where $B_\nu(T)$ is a blackbody intensity spectrum. The same relation holds for the
Stokes parameters $Q$ and $U$.

We use the brightness temperature as an intermediate step between the
intensity and thermodynamic temperature. The brightness-thermodynamic
temperature conversion factor $g_\nu$ (Eq.~\ref{eq:Fnu}) with bandpass
averaging is then given by
\begin{equation}
 \left<g_\nu\right> = \frac{2k_{B}\nu^2/c^2}{\left<\frac{\partial B_\nu(T)}{\partial T}\right>}
  = \frac{\nu^2}{\left<\nu^2\frac{e^xx^2}{(e^x-1)^2}\right>}
~.
\end{equation}
There is no averaging for $\nu^2$ in the numerator because it is the
intensity that is averaged, and conversion from the intensity to
brightness temperature is merely an intermediate step.
Using these quantities, our sky model
(Eq.~(\ref{eq:QUmodel})) is now expressed as
\begin{eqnarray}
 \nonumber
 [Q,U]_\nu(\hat{n}) &=& \mbox{CMB}(\hat{n})
  +\left<g_\nu\right> \frac{\left<\nu^2D_{\nu}(p^I(\hat{n}))\right>}{\nu^2}[Q^f,U^f]_{\nu_\ast}(\hat{n})
  +N_{\nu}(\hat{n})\\
 &=&   \mbox{CMB}(\hat{n})
  +\frac{\left<\nu^2D_{\nu}(p^I(\hat{n}))\right>}{\left<\nu^2\frac{e^xx^2}{(e^x-1)^2}\right>}[Q^f,U^f]_{\nu_\ast}(\hat{n})
  +N_{\nu}(\hat{n})\,.
\label{eq:QUmodel_av}
\end{eqnarray}
As $[Q^f,U^f]_{\nu_\ast}$ is an auxiliary variable, there is no need to
bandpass-average it. The average is done only for variables depending on
$\nu$. Therefore, to incorporate bandpass averaging, all we need is to
replace $g_\nu D_\nu$ with $\langle g_\nu\rangle\langle\nu^2D_\nu\rangle/\nu^2$.

Incorporating bandpass averaging in the foreground removal algorithm is
important because it changes the effective frequency at which intensity
of foregrounds should be evaluated. Using an incorrect bandpass leads to
a bias in $r$. How precisely we should know
the bandpass needs to be determined by the requirement for science,
e.g., a target value of $r$, coupled with a careful analysis including
foreground removal and systematic errors in our knowledge of the
bandpass. This is work in progress. For the rest of this paper we shall
ignore bandpass averaging in the simulations and leave detailed study of
its effect for future work.

\section{Implementation}
\label{sec:implementation}
\subsection{Cleaning one map}
Our likelihood for a cleaned CMB map $\vec{x}$
(Eq.~(\ref{eq:cleanedmap})) is
\begin{equation}
-2\ln{\cal L} = \vec{x}(\bar{p}^I)^{\rm T}{\bm C}^{-1}(r,s,\bar{p}^I)\vec{x}(\bar{p}^I)+\ln|2\pi{\bm C}|,
 \label{eq:(cal L)}
\end{equation}
where the covariance matrix ${\bm{C}}$ is given by \cite{2011ApJ...737...78K}
\begin{equation}
 {\bm C}(r,s,\bar{p}^I) = {\bm C}^{\rm tens}(r)+{\bm C}^{\rm scal}(s) +
  \frac{{\bm N}_{\nu_{\rm CMB}}+\sum^{N+1}_{j=1}\alpha_j^2 {\bm
  N}_{\nu_j}}{\left(1+\sum_{j=1}^{N+1}\alpha_j\right)^2} + {\bm E}_{\rm diag}~,
\label{eq:covariancematrix}
\end{equation}
and ${\bm C}^{\rm tens}$ and ${\bm C}^{\rm scal}$ are the tensor and
scalar CMB signal covariance matrices (including beam
smearing, pixel window function, and any other extra smoothing/filtering applied
to data), respectively,
calculated from theoretical power spectra following Appendix A of
\cite{2011ApJ...737...78K}. Noise covariance matrices $\bm{N}_\nu$
and $\bm{E}_{\rm diag}$ will be explained after Eq.~(\ref{eq:artificialnoise}).

Theoretical power spectra are given by
\begin{eqnarray}
 C_\ell^{\rm EE} &=& rc_\ell^{\rm tens,EE}+sc_\ell^{\rm scal,EE} ~,\\
 C_\ell^{\rm BB} &=& rc_\ell^{\rm tens,BB}+sc_\ell^{\rm lens,BB}~,
\end{eqnarray}
where $c_\ell^{\rm tens,EE}$ and $c_\ell^{\rm tens,BB}$ are the E-
and B-mode polarization power spectra from the tensor mode with $r=1$,
respectively, and $c_\ell^{\rm scal,EE}$ and $c_\ell^{\rm lens,BB}$
are the E mode from the scalar mode and the B mode from the
gravitational-lensing effect, respectively.

The parameter $s$ rescales the amplitude of the scalar perturbation, and
the fiducial value is $s=1$. In \cite{2011ApJ...737...78K}, we found
that marginalization over the scalar amplitude $s$ was able to
remove a chance correlation between the template coefficients and the
tensor-to-scalar ratio $r$. 

To compute the fiducial theoretical power spectra, we use the Planck
2015 cosmological parameters for ``TT$+$LowP$+$lensing''
\cite{2016A&A...594A..13P}: $\Omega_bh^2=0.02226$, $\Omega_ch^2=0.1186$,
$h=0.6781$, $\tau=0.066$, $A_s=2.137\times 10^{-9}$, and $n_s=0.9677$.

In this paper, we work with low-resolution maps at a Healpix
\cite{2005ApJ...622..759G} resolution of $N_{\rm side}=4$. At this
resolution and with very low instrumental noise we assume in this paper
(anticipating future space missions such as LiteBIRD
\cite{2016SPIE.9904E..0XI}), the signal covariance matrix becomes
ill-conditioned due to a strong scalar E-mode signal which is not
band-limited. To solve this numerical issue, we follow the procedure
described in Appendix A of \cite{2007ApJ...656..641E}; specifically, we
apply a Gaussian smoothing to a signal-plus-noise map with 2200~arcmin (FWHM), which is 2.5 times the pixel size at $N_{\rm side}=4$,
and resample the smoothed map to $N_{\rm side} = 4$. As the smoothed map at $N_{\rm side} =
4$ is dominated by the scalar E-mode signal at all angular scales supported by
the map resolution, the covariance matrix of this map
is singular. In order to regularize the covariance matrix,
we add an artificial, homogeneous white noise of 0.2~$\mu{\rm K}$ arcmin
to the smoothed map, such that the map becomes noise dominated at the
Nyquist frequency of the map:
\begin{equation}
 \vec{x} = \frac{[Q,U]_{\nu_{\rm CMB}}(\hat{n}) +
  \sum_{j=1}^{N+1}\alpha_j[Q,U]_{\nu_j}(\hat{n})}{1+\sum_{j=1}^{N+1}\alpha_j}
  + \mbox{artificial noise}(\hat{n})~.
\label{eq:artificialnoise}
\end{equation}
The noise covariance matrices ${\bm N_{\nu_{\rm
CMB}}}$ and ${\bm N_{\nu_j}}$ in Eq.~(\ref{eq:covariancematrix}) become
non-diagonal because of smoothing, and ${\bm E}_{\rm diag}$ is a
diagonal matrix for the artificial noise used to regularize the
covariance matrix.

We now wish to estimate $\bar{p}^I$, $r$, and $s$ from the likelihood
given in Eq.~(\ref{eq:(cal L)}). We follow the following procedure:
\begin{itemize}
 \item[1.] {\bf Initialization:} Set initial values for $s$, $r$, and
	   $\bar{p}^I$;
 \item[2.] {\bf Iteration step I:} Minimize the $\chi^2$ part of the
	   likelihood (the first term in Eq.~(\ref{eq:(cal L)}))\footnote{One may wonder why we do not maximize the
	   likelihood for $\bar{p}^I$, $r$, and $s$ simultaneously using
	   Eq.~(\ref{eq:(cal L)}). We provide a reason for this in
	   Sect.~\ref{sec:bayesian}.}
\begin{equation}
\bar{p}^I=\underset{\bar{p}}{\arg\min}~\vec{x}(\bar{p}^I)^{\rm T}{\bm
 C}^{-1}(r,s,\bar{p}^I)\vec{x}(\bar{p}^I),
 \label{eq:chi2min}
\end{equation}
with $s$ and $r$ fixed. During minimization, the covariance matrix
 ${\bm C}$ and its inverse ${\bm C}^{-1}$ are updated at each
	   evaluation of $\bar{p}^I$.
 \item[3.] {\bf Iteration step II:} Maximize the likelihood given in
	   Eq.~(\ref{eq:(cal L)}) for $s$ and $r$ with the foreground parameters
	   $\bar{p}^I$ fixed to the values found during the iteration
	   step I. During maximization, ${\bm C}$, ${\bm C}^{-1}$, and
	   the determinant $|{\bm C}|$ are updated at each evaluation of
	   $s$ and $r$.
 \item[4.] Return to step 2 until a convergence criterion is satisfied.
\end{itemize}
We find that ten iterations are typically sufficient to meet the
convergence criterion we set as $|r_{\rm new}-r_{\rm old}|< 10^{-3}r_{\rm
old}$.  To maximize (minimize) the functions we have used the {\it nlopt}
library\footnote{https://nlopt.readthedocs.io/en/latest/}.

Finally, we include the mask in the covariance matrix following
\cite{2007ApJS..170..335P} (see also Appendix~\ref{sec:mask}). We use
the ``P06 mask'' defined by the WMAP polarization analysis, which
retains 73\% of sky for the analysis. We degrade the original P06 mask
to $N_{\rm side}=4$. 

For $N_{\rm side}=4$ simulations, analyzing $1000$ maps roughly takes $1000$
seconds on a laptop (4 cores, 16GB memory).  The most
time-consuming part is matrix inversion, whose computational cost scales
as the number of matrix elements cubed; thus, the computational cost
scales as $\propto N_{\rm side}^6$.

\subsection{Cleaning multiple maps}
\label{sec:multiclean}
We may use the Delta-map method to clean two (or more) maps at two (or more)
CMB frequencies. Here we consider a case in which we clean two
maps at different CMB frequencies (e.g., $100$ GHz and $140$ GHz) using
lower- and higher-frequency maps as the common template maps for
cleaning. Having two frequency channels cleaned separately (instead of
combining them into one cleaned map) is useful for null tests, as we
describe in Sect.~\ref{sec:null}.

In this case, the map vector becomes
\begin{equation}
\vec{x}^\prime = 
\begin{pmatrix}
\vec{x}_{\nu_{\rm CMB1}} \\
\vec{x}_{\nu_{\rm CMB2}} 
\end{pmatrix}
~,
\end{equation}
where $\vec{x}_{\nu_{\rm CMB1,2}}$ are cleaned CMB maps at
frequencies $\nu_{\rm CMB1}$ and $\nu_{\rm CMB2}$, given by
\begin{equation}
 \vec{x}_{\nu_{\rm CMB1,2}} = \frac{[Q,U]_{\nu_{\rm CMB1,2}}(\hat{n}) +
  \sum_{j=1}^{N+1}\alpha_j^{\nu_{\rm CMB1,2}}[Q,U]_{\nu_j}(\hat{n})}{1+\sum_{j=1}^{N+1}\alpha_j^{\nu_{\rm CMB1,2}}}
  + \mbox{artificial noise}_{1,2}(\hat{n})~.
\end{equation}
The noise covariance matrix for this map vector is given by
\begin{equation}
 \bm{N}(\hat{n},\hat{n}^\prime)=
\begin{pmatrix}
\bm{n}_{Q_1Q_1} & \bm{n}_{Q_1U_1} & \bm{n}_{Q_1Q_2} & \bm{n}_{Q_1U_2} \\
\bm{n}_{U_1Q_1} & \bm{n}_{U_1U_1} & \bm{n}_{U_1Q_2} & \bm{n}_{U_1U_2} \\
\bm{n}_{Q_2Q_1} & \bm{n}_{Q_2U_1} & \bm{n}_{Q_2Q_2} & \bm{n}_{Q_2U_2} \\
\bm{n}_{U_2Q_1} & \bm{n}_{U_2U_1} & \bm{n}_{U_2Q_2} & \bm{n}_{U_2U_2} \\
\end{pmatrix},
\end{equation}
where the components of the noise covariance matrix are, for example,
\begin{eqnarray}
 \bm{n}_{Q1Q1} &=& \frac{\bm{n}^{QQ}_{\nu_{\rm CMB1}}
+\sum_{j=1}^{N+1}\left(\alpha_j^{\nu_{\rm CMB1}}\right)^2 \bm{n}^{QQ}_{\nu_j}
}{\left(1+\sum_{j=1}^{N+1} \alpha_j^{\nu_{\rm CMB1}}\right)^2},\\
\bm{n}_{Q1U2}&=& \frac{\sum_{j=1}^{N+1}\alpha_{j}^{\nu_{\rm CMB1}}\alpha_{j}^{\nu_{\rm CMB2}}\bm{n}^{QU}_{\nu_j}
}
{
\left(1+\sum_{j=1}^{N+1} \alpha_j^{\nu_{\rm CMB1}}\right)
\left(1+\sum_{j=1}^{N+1} \alpha_j^{\nu_{\rm CMB2}}\right)
}.
\end{eqnarray}
Here $\bm{n}^{QQ}_{\nu}$, $\bm{n}^{QU}_{\nu}$, and $\bm{n}^{UU}_{\nu}$
are the noise covariance matrices for the Stokes $Q$ and $U$ maps at
frequency $\nu$.
The signal covariance is given more simply by
\begin{equation}
 \bm{S}(\hat{n},\hat{n}^\prime) = 
\begin{pmatrix}
\bm{s}_{QQ} & \bm{s}_{QU} & \bm{s}_{QQ} & \bm{s}_{QU} \\
\bm{s}_{UQ} & \bm{s}_{UU} & \bm{s}_{UQ} & \bm{s}_{UU} \\
\bm{s}_{QQ} & \bm{s}_{QU} & \bm{s}_{QQ} & \bm{s}_{QU} \\
\bm{s}_{UQ} & \bm{s}_{UU} & \bm{s}_{UQ} & \bm{s}_{UU} \\
\end{pmatrix},
\end{equation}
where $\bm{s}_{QQ},\bm{s}_{QU},\bm{s}_{UQ}$ and $\bm{s}_{UU}$ are the
signal covariance matrices for the Stokes $Q$ and $U$ of the CMB. Here,
the same signal matrix elements appear for different
frequency channels; however, it should be understood that they are
different in practice due to different beam smearing and filtering
depending on channels. We have not written this effect explicitly for
simplicity throughout this paper.

\subsection{Test}
\label{sec:algorithm-test} Before moving on to simulations with
realistic foreground models, we test the algorithm
using simplified ones. Specifically, we generate foreground maps
matching exactly the form assumed in Eq.~(\ref{eq:simple_delta}),
i.e., power-law emission expanded and truncated at first order in
$\delta\beta$. For $Q$ and
$U$ maps and power-law indices of synchrotron and dust, we use the same
maps as in \cite{2011ApJ...737...78K}, in which the mean synchrotron and
dust indices are $\bar{\beta}_{\rm
s}=-3.00$ and $\bar{\beta}_{\rm d}=1.647$, respectively.

We add scalar and tensor modes of CMB with a tensor-to-scalar ratio of
$r=0.001$. We do not add instrumental noise but add artificial noise to
regularize the covariance matrix. We then apply the Delta-map method to
estimate $r$, $\bar{\beta}_{\rm d}$, and $\bar{\beta}_{\rm s}$.

\begin{figure}[t]
 \centering\includegraphics[width=1.0\linewidth]{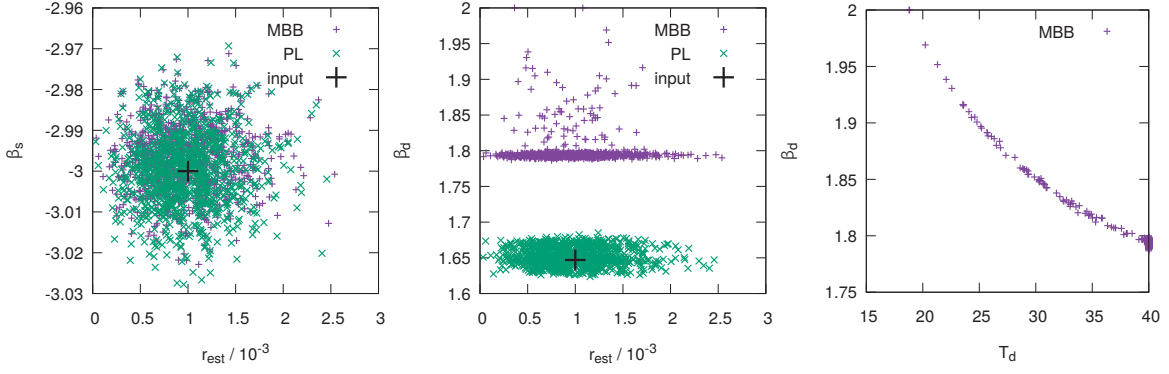}    
 \caption{Values of $r$ and the mean foreground parameters obtained from 1000
   random realizations of the CMB. No instrumental noise is
   included. The simulation also includes power-law 
   synchrotron and power-law dust with spatially-varying indices, and the mean
   indices are $\bar{\beta}_{\rm s}=-3.0$ and $\bar{\beta}_{\rm
   d}=1.647$. While we always use the Delta-map method for a power-law
   spectrum (Sect.~\ref{sec:2.2.1}) for synchrotron, we compare two
   Delta-map methods for dust: power-law and MBB (Sect.~\ref{sec:MBB}). The
   input map for dust is always a power-law (only in this test). The
   green (denoted as ``PL'') and
   purple (denoted as ``MBB'') points show the results from the
   Delta-map method for power-law dust and MBB dust, respectively. We use flat priors with $r=[0,1]$, $\bar{\beta}_{\rm s}=[-4,-2]$, $\bar{\beta}_{\rm d}=[1,2]$, and
   $T_{\rm d}=[10~{\rm K},40~{\rm K}]$.}  \label{fig:fig1}
  \end{figure}

For this test we use CMB$+$foreground maps generated at 40, 60, 140, 230, and
340~GHz, and use the lowest two and highest two frequencies to clean a
map at 140~GHz. We do not apply bandpass averaging.
Fig.~\ref{fig:fig1} shows the values of $r$, $\bar{\beta}_{\rm s}$,
and $\bar{\beta}_{\rm d}$ obtained from 1000 random realizations of the
CMB. We successfully recover the input values.

Next, we apply the Delta-map method for a MBB for
dust (Sect.~\ref{sec:MBB}), while the simulated maps always have a power-law
dust (only in this section). For this case we generate another map generated at 400~GHz so that
we have the highest three frequencies for dust.
We find that we recover the input values of $r$ and
$\bar{\beta}_{\rm s}$. This is because, in the limited frequency
range considered here, a MBB can mimic a power law with a spectral index of
$1.647$ by appropriately tuning $\bar{\beta}_{\rm d}$ and
$\bar{T}_{\rm d}$. (MBB becomes a power law in the Rayleigh-Jeans limit,
i.e., $k_BT_{\rm d}/h\nu \to \infty$.) The clear correlation seen in the
right panel of Fig.~\ref{fig:fig1} supports this interpretation. 

\section{Bayesian derivation of the Delta-map method}
\label{sec:bayesian}
\subsection{Likelihood}
So far we have formulated the Delta-map method in a ``bottom-up
approach''; namely, we started with a plausible form of a foreground-cleaned
CMB map given in Eq.~(\ref{eq:cleanedmap}) and constructed a likelihood given in Eq.~(\ref{eq:(cal L)}). In this section we formulate the
Delta-map method in a ``top-down approach'' by starting with the
likelihood of data $\vec{m}$ given general sky signal $\vec{s}$, mixing
matrix $\bm{D}$, and noise covariance matrix $\bm{N}$:
\begin{equation}
 -2\ln{\cal L}(\vec{m}|\vec{s},\bm{D})=\left(\vec{m}-\bm{D}\vec{s}\right)^{\rm
  T}\bm{N}^{-1}\left(\vec{m}-\bm{D}\vec{s}\right)
  +\ln|2\pi\bm{N}|  \,.
  \label{eq:bayesianlikelihood}
\end{equation}
We have concatenated the Stokes parameter data at all pixels and frequencies as 
\begin{equation}
 \vec{m}\equiv\left(\{[Q,U]_{\nu_1}(\hat
	       n_i)]\},\cdots,\{[Q,U]_{\nu_{N_{\rm freq}}}(\hat
	       n_{i})\}\right)^{\rm T}\,.
\end{equation}
We similarly concatenate the noise covariance matrix $\bm{N}$, which now
becomes a $2N_{\rm pix}N_{\rm freq}$-by-$2N_{\rm pix}N_{\rm freq}$
matrix. (The factor of 2 is for $Q$ and $U$.)
Here, $N_{\rm pix}$ is the number of pixels and $N_{\rm freq}$ is the
number of frequency channels, and $\hat n_i$ is a
sky direction unit vector toward $i$th pixel.
We have also defined the sky signal vector as
\begin{equation}
 \vec{s}\equiv \left(\{\mbox{CMB}(\hat n_i)\},\{\mbox{Dust}(\hat
		n_i)\},\{\mbox{Synch}(\hat n_i)\},\cdots\right)^{\rm T}\,,
\end{equation}
whose number of elements is $2N_{\rm pix}(N_{\rm temp}+1)$, where $N_{\rm
temp}$ is the number of foreground (``template'') components. When we
have synchrotron and dust, $N_{\rm temp}=2$.

The number of rows of the mixing matrix $\bm{D}$ is $2N_{\rm pix}N_{\rm
freq}$, and the number of columns is $2N_{\rm pix}(N_{\rm temp}+1)$.
We may write $\bm{D}$ using $2N_{\rm pix}N_{\rm freq}$-by-$2N_{\rm pix}$ sub-matrices as
\begin{equation}
 {\bm D}=
   \begin{pmatrix}
\bm{D}^{\rm CMB}&\bm{D}^{\rm dust}&\bm{D}^{\rm synch}&\cdots
   \end{pmatrix}\,,
\end{equation}
where
\begin{equation}
 {\bm D}^{\rm CMB} \equiv
 \begin{pmatrix}
  1 & 0 & \cdots & 0\\
  0 & 1 & \cdots & 0\\
  \vdots & \vdots & \ddots & \vdots \\
  0 & 0 & \cdots & 1 \\
  1 & 0 & \cdots & 0\\
  0 & 1 & \cdots & 0\\
  \vdots & \vdots & \ddots & \vdots \\
  0 & 0 & \cdots & 1\\
    \vdots & \vdots & \vdots & \vdots
 \end{pmatrix}\,,
\end{equation}
\begin{equation}
 {\bm D}^{\rm dust} \equiv
  \begin{pmatrix}
   g_{\nu_1} D^{\rm dust}_{\nu_1}(\beta_{{\rm d},1},T_{{\rm
 d},1},\cdots) & 0 & \cdots & 0\\
   0 & g_{\nu_1} D^{\rm dust}_{\nu_1}(\beta_{{\rm d},2},T_{{\rm
 d},2},\cdots)  & \cdots & 0\\
  \vdots & \vdots & \ddots & \vdots \\
  0 & 0 & \cdots & g_{\nu_1} D^{\rm dust}_{\nu_1}(\beta_{{\rm d},N_{\rm pix}},T_{{\rm
 d},N_{\rm pix}},\cdots) \\
   g_{\nu_2} D^{\rm dust}_{\nu_2}(\beta_{{\rm d},1},T_{{\rm
 d},1},\cdots) & 0 & \cdots & 0\\
   0 & g_{\nu_2} D^{\rm dust}_{\nu_2}(\beta_{{\rm d},2},T_{{\rm
 d},2},\cdots)  & \cdots & 0\\
  \vdots & \vdots & \ddots & \vdots \\
  0 & 0 & \cdots & g_{\nu_2} D^{\rm dust}_{\nu_2}(\beta_{{\rm d},N_{\rm pix}},T_{{\rm
   d},N_{\rm pix}},\cdots) \\
       \vdots & \vdots & \vdots & \vdots
  \end{pmatrix}\,,
\end{equation}
and similarly for $\bm{D}^{\rm synch}$. Here, $\beta_{{\rm d},i}\equiv
\beta_{{\rm d}}(\hat n_i)$ etc., and each block in $\bm{D}^{\rm dust}$
should include the entries for both $Q$ and $U$, i.e., each block is a
$2N_{\rm pix}$-by-$2N_{\rm pix}$ matrix.

The effect we have not written explicitly here is the effect of a
gain calibration, beam smearing, pixel window function, or any extra
smoothing/filtering applied to data. These effects will modify all
elements of the mixing matrix; e.g., ``$1$'' in the CMB matrix becomes a product of the
calibration factor, beam transfer and pixel window function operators, etc.

The next step is to expand the mixing matrix to first order in spatial
variations of the spectral parameters. We redefine the signal vectors as
\begin{align}
\nonumber
 \vec{s}\to \Big(\{\mbox{CMB}(\hat n_i)\},&\{\mbox{Dust}(\hat
  n_i)\},\{\mbox{Dust}(\hat n_i)\delta\beta_{{\rm
  d},i}\},\{\mbox{Dust}(\hat n_i)\delta T_{{\rm
 d},i}\},\cdots, \\
 & \{\mbox{Synch}(\hat n_i)\}, \{\mbox{Synch}(\hat
  n_i)\delta\beta_{{\rm s},i}\},\{\mbox{Synch}(\hat n_i)\delta C_{{\rm
 s},i}\},\cdots\Big)^{\rm T}\,,
\end{align}
as well as the mixing matrix as
\begin{equation}
 {\bm D}\to
   \begin{pmatrix}
\bm{D}^{\rm CMB}&\bm{D}^{\rm dust}(\bar{p}^I)&\bm{D}^{\rm
    dust}_{,1}&\cdots&\bm{D}^{\rm dust}_{,N}&\bm{D}^{\rm synch}(\bar{p}^J)&\bm{D}^{\rm
    synch}_{,1}&\cdots&\bm{D}^{\rm synch}_{,M}&
    \cdots
   \end{pmatrix}\,,
\end{equation}
where dust and synchrotron have $N$ ($I=1,\cdots,N$) and $M$
($J=1,\cdots,M$) spectral parameters,
respectively. The subscript ``$,I$'' denotes derivatives with respect to the
$I$th foreground parameter (e.g., $\beta_{\rm d}$ and $T_{\rm d}$ for
dust and  $\beta_{\rm s}$ and $C_{\rm s}$ for synchrotron).
All elements in this new mixing matrix are evaluated at the mean
spectral parameters. The form of the likelihood given in
Eq.~(\ref{eq:bayesianlikelihood}) remains unchanged.

Specifically,
 \begin{align}
  \nonumber
  &
  \begin{pmatrix}
 \bm{D}^{\rm dust}(\bar{p}^I)&\bm{D}^{\rm
   dust}_{,1}&\cdots&\bm{D}^{\rm dust}_{,N}
  \end{pmatrix}\\
 =&
    \begin{pmatrix}
     g_{\nu_1} D^{\rm dust}_{\nu_1}(\bar{p}^I)\bm{I} &
     g_{\nu_1} D^{\rm dust}_{\nu_1,1}\bm{I}&\cdots& g_{\nu_1} D^{\rm dust}_{\nu_1,N}\bm{I}
     \\
      g_{\nu_2} D^{\rm dust}_{\nu_2}(\bar{p}^I)\bm{I} &
     g_{\nu_2} D^{\rm dust}_{\nu_2,1}\bm{I}&\cdots& g_{\nu_2} D^{\rm dust}_{\nu_2,N}\bm{I}
     \\
     \vdots & \vdots &\ddots &\vdots \\
     g_{\nu_{\rm Nfreq}} D^{\rm dust}_{\nu_{N_{\rm freq}}}(\bar{p}^I)\bm{I} &
     g_{\nu_{N_{\rm freq}}} D^{\rm dust}_{\nu_{N_{\rm freq}},1}\bm{I}&\cdots
     &g_{\nu_{N_{\rm freq}}} D^{\rm dust}_{\nu_{N_{\rm freq}},N}\bm{I}
  \end{pmatrix}\,,
\end{align}
and similarly for synchrotron etc. Here, $\bm{I}$ is a $2N_{\rm
pix}$-by-$2N_{\rm pix}$ identity matrix.

When we have only CMB, dust and synchrotron in the sky, the number of
elements of $\vec{s}$ is $2N_{\rm pix}(N+M+3)$ and $\bm{D}$ becomes a
$2N_{\rm pix}N_{\rm freq}$-by-$2N_{\rm pix}(N+M+3)$ matrix. When
$N_{\rm freq}=N+M+3$, $\bm{D}$ becomes a square matrix and
invertible. This is a special case in which the number of frequency
channels is just enough to solve for a cleaned CMB map.

Now that the mixing matrix $\bm{D}$ does not depend on pixels,
Eq.~(\ref{eq:bayesianlikelihood}) can be written trivially in pixel, spherical
harmonics, or wavelet space. This is an important advantage of
linearization used by the Delta-map method.

\subsection{Delta-map as the maximum likelihood solution}
Maximizing the likelihood with respect to the CMB
signal, we obtain \cite{2009MNRAS.392..216S}
\begin{equation}
 \mbox{CMB}^{\rm ML}(\hat n)
  =\left[(\bm{D}^{\rm T}\bm{N}^{-1}\bm{D})^{-1}\bm{D}^{\rm
    T}\bm{N}^{-1}\vec{m}   \right]_{\rm CMB}\,,
\label{eq:exactML}
\end{equation}
where the subscript ``CMB'' indicates that we take the elements
relevant to the CMB, and the superscript ``ML'' stands for the maximum
likelihood value. We can evaluate this in pixel, spherical harmonics, or
wavelet space.

When the number of frequency channels is just enough to solve for a
cleaned CMB map, $\bm{D}$ is invertible and we obtain
$\mbox{CMB}^{\rm ML}=\left[\bm{D}^{-1}\vec{m}\right]_{\rm CMB}$.
For example, when we have a spatially uniform
power-law component as foreground, the solution for a cleaned
CMB map from two maps at two frequencies is
\begin{equation}
 \mbox{CMB}^{\rm ML}(\hat n)
  = \frac{g_{\nu_2}D_{\nu_2}m_{\nu_1}(\hat
  n)-g_{\nu_1}D_{\nu_1}m_{\nu_2}(\hat
  n)}{g_{\nu_2}D_{\nu_2}-g_{\nu_1}D_{\nu_1}}\,,
  \label{eq:simplestML}
\end{equation}
where $D_\nu=(\nu/\nu_*)^{\bar\beta}$. Note that
$\nu_*$ cancels. If we identify $\nu_2$ with the reference frequency $\nu_*$,
the expression further simplifies to
$\mbox{CMB}^{\rm ML}(\hat n)
  = [m_{\nu_1}(\hat n)-(g_{\nu_1}/g_{\nu_*})D_{\nu_1}m_{\nu_2}(\hat
  n)]/[1-(g_{\nu_1}/g_{\nu_*})D_{\nu_1}]$, which coincides with the usual
  internal template cleaning solution with
  $\alpha=(g_{\nu_1}/g_{\nu_*})D_{\nu_1}$ (see, e.g., Eq.~(6) of \cite{2011ApJ...737...78K}).
We can obtain the same result from Eq.~(\ref{eq:cleanedmap}) with
$\alpha$ given by $-\bm{A}^{-1}\vec{d}_{\nu_{\rm CMB}}$ and
$\nu_1=\nu_{\rm CMB}$.

When we have a spatially-varying power-law component as
foreground, the solution for a cleaned
CMB map from three maps at three frequencies is
 \begin{equation}
 \mbox{CMB}^{\rm ML}(\hat n)
 =\frac{\xi_{12}m_{\nu_3}(\hat n)+\xi_{31}m_{\nu_2}(\hat n)
  - \xi_{23}m_{\nu_1}(\hat
  n)}{\xi_{12}+\xi_{31}-\xi_{23}}\,,
 \end{equation}
 where
 $\xi_{ij}\equiv g_{\nu_i}g_{\nu_j}\left(D_{\nu_i,\beta}D_{\nu_j}
 -D_{\nu_i}D_{\nu_j,\beta}\right)=
 g_{\nu_i}g_{\nu_j}\left(\nu_i\nu_j/\nu_*^2\right)^{\bar\beta}\ln(\nu_i/\nu_j)$.
 The numerator evaluates to
  \begin{align}
\nonumber
& g_{\nu_1}g_{\nu_2}(\nu_1\nu_2)^{\bar\beta}\ln\left(\frac{\nu_1}{\nu_2}\right)
   m_{\nu_3}(\hat n)\\
+&
 g_{\nu_3}g_{\nu_1}(\nu_3\nu_1)^{\bar\beta}\ln\left(\frac{\nu_3}{\nu_1}\right)m_{\nu_2}(\hat
   n)
     -
  g_{\nu_2}g_{\nu_3}(\nu_2\nu_3)^{\bar\beta}
  \ln\left(\frac{\nu_2}{\nu_3}\right)m_{\nu_1}(\hat n)\,,
 \end{align}
 where we have canceled $\nu^*$ with that in the denominator.
 We find that this result reproduces the numerator of Eq.~(\ref{eq:deltamapresultpowerlaw}), $[Q,U]_{\nu_{\rm
 CMB}}+\sum_{j=1}^2[Q,U]_{\nu_j}$, up to a multiplicative  factor if we
 identify $\nu_3$ with $\nu_{\rm CMB}$. 

We thus conclude that Eq.~(\ref{eq:cleanedmap}) gives the maximum likelihood
solution for a cleaned CMB map to first order in perturbation, provided
that the number of frequency channels is just enough to solve for a
cleaned CMB map. We need to use Eq.~(\ref{eq:exactML}) when we have more
channels, which would give a less noisy CMB map.

\subsection{Should we cancel CMB in a template map?}
SEVEM (see Appendix A.6 of \cite{2008A&A...491..597L}) is an internal
template method used by the Planck collaboration. In this method one
constructs a template foreground map by first taking difference
between two frequency channels (so that CMB cancels) and fits this
difference map to a CMB channel and subtracts the
foreground. Specifically, when we have one foreground component and
three frequency channels, SEVEM constructs
\begin{equation}
 \mbox{CMB}^{\rm SEVEM}(\hat n)=m_{\nu_1}(\hat n)-\alpha\left[m_{\nu_2}(\hat
  n)-m_{\nu_3}(\hat n)\right]\,,
\end{equation}
where $\alpha$ is a fitting coefficient. (For this case, $\nu_{\rm CMB}=\nu_1$.)
When the foreground has a spatially uniform spectrum, the coefficient is
$\alpha=g_{\nu_1}D_{\nu_1}/(g_{\nu_2}D_{\nu_2}-g_{\nu_3}D_{\nu_3})$,
which gives an unbiased estimate of CMB.

We find that this is not the maximum likelihood solution given in
Eq.~(\ref{eq:exactML}). Assuming that the noise covariance matrix is
diagonal and homogeneous over pixels, $\bm{N}={\rm
diag}(\sigma_{\nu_1}^2\bm{I},\sigma_{\nu_2}^2\bm{I},\sigma_{\nu_3}^2\bm{I})$,
where $\bm{I}$ is a $2N_{\rm pix}$-by-$2N_{\rm pix}$ identity matrix,
Eq.~(\ref{eq:exactML}) becomes
\begin{equation}
 \mbox{CMB}^{\rm ML}(\hat n)
  =\frac{(g_{\nu_1}D_{\nu_1}-g_{\nu_2}D_{\nu_2})
  \left[g_{\nu_1}D_{\nu_1}m_{\nu_2}(\hat
   n)-g_{\nu_2}D_{\nu_2}m_{\nu_1}(\hat
   n)\right]\sigma_{\nu_3}^2+(\mbox{2 perm.})}
  {(g_{\nu_1}D_{\nu_1}-g_{\nu_2}D_{\nu_2})^2\sigma_{\nu_3}^2+(\mbox{2 perm.})}\,,
\end{equation}
where ``2 perm.'' indicates two permutation terms.
While this may look complicated, there is a clear interpretation. To see
this, let us define
\begin{equation}
 \mbox{CMB}^{\rm ML,2band}_{ij}(\hat n)\equiv \frac{g_{\nu_i}D_{\nu_i}m_{\nu_j}(\hat
  n)-g_{\nu_j}D_{\nu_j}m_{\nu_i}(\hat
  n)}{g_{\nu_i}D_{\nu_i}-g_{\nu_j}D_{\nu_j}}\,,
\end{equation}
which is the Maximum likelihood solution when we have only two frequency
channels $\nu_i$ and $\nu_j$ (Eq.~(\ref{eq:simplestML})). We find
\begin{equation}
 \mbox{CMB}^{\rm ML}(\hat n)
  =\frac{(g_{\nu_1}D_{\nu_1}-g_{\nu_2}D_{\nu_2})^2
  \mbox{CMB}^{\rm ML,2band}_{12}(\hat n)\sigma_{\nu_3}^2+(\mbox{2 perm.})}
  {(g_{\nu_1}D_{\nu_1}-g_{\nu_2}D_{\nu_2})^2\sigma_{\nu_3}^2+(\mbox{2 perm.})}\,.
\end{equation}
Therefore, the maximum likelihood solution is equal to three cleaned CMB
maps combined optimally, but is not equal to the SEVEM construction.

\subsection{Posterior of the parameters}
\label{sec:posterior}
The next task is to derive a posterior distribution of the mean spectral
parameters $\bar p^I$ and the cosmological parameters such as the
tensor-to-scalar ratio $r$. We start by using Bayes's theorem to relate
the posterior distribution to the likelihood given in
Eq.~(\ref{eq:bayesianlikelihood}) as
\begin{equation}
 P(\bar p^I,\vec{s},\bm{S}|\vec{m}) = \frac{{\cal
  L}(\vec{m}|\vec{s},\bm{D})P(\bm{D},\vec{s},\bm{S})}{P(\vec{m})}\,,
  \label{eq:bayestheorem}
\end{equation}
where $\bm{S}$ is a covariance matrix of the sky signal 
$\vec{s}=\vec{s}_{\rm CMB}+\vec{s}_f$ (CMB $\vec{s}_{\rm CMB}$ and 
foregrounds $\vec{s}_f$), $P(\bm{D},\vec{s},\bm{S})$ is the prior
distribution, and $P(\vec{m})$ is Bayes's
(normalization) factor. Note that $\bar p^I$ is a vector of the parameters that
determine the mixing matrix $\bm{D}$. We shall assume a flat prior on
$\bm{D}$, i.e., $P(\bm{D},\vec{s},\bm{S})\propto P(\vec{s},\bm{S})$.

We first marginalize Eq.~(\ref{eq:bayestheorem}) over the
CMB signal. To this end we assume a Gaussian probability density
distribution for the CMB signal part of the prior:
$-2\ln P(\vec{s},\bm{S})=\vec{s}_{\rm CMB}(\bm{S}_0^{\rm
CMB})^{-1}\vec{s}_{\rm CMB}+\ln|2\pi\bm{S}_0^{\rm CMB}|+(\mbox{other
terms})$, where $\bm{S}_0^{\rm CMB}$ is a $2N_{\rm pix}$-by-$2N_{\rm
pix}$ CMB signal covariance matrix
with no beam smearing, pixel window function, extra filtering, or calibration factor.

Integrating both sides of Eq.~(\ref{eq:bayestheorem}) over $\vec{s}_{\rm CMB}$, we obtain
\begin{eqnarray}
\nonumber
 -2\ln P(\bar p^I,\vec{s}_{f},\bm{S}|\vec{m}) &=&
\left(\vec{m}-\tilde{\bm{D}}\vec{s}_{f}\right)^{\rm
  T}\left(\bm{S}^{\rm CMB}+\bm{N}\right)^{-1}\left(\vec{m}-\tilde{\bm{D}}\vec{s}_{f}\right)\\
& &     +\ln\left|2\pi(\bm{S}^{\rm CMB}+\bm{N})\right|
-2\ln P(\vec{s}_{f},\bm{S}^{f})+\mbox{constant}  \,,
     \label{eq:tobemaximized}
\end{eqnarray}
where $\bm{S}^{\rm CMB}=\bm{D}^{\rm CMB}\bm{S}^{\rm CMB}_0(\bm{D}^{\rm
CMB})^{\rm T}$, $\vec{s}_{f}$ is the foreground sky signal vector with
$P(\vec{s}_{f},\bm{S}^f)$ being its prior, and
$\tilde{\bm{D}}$ is a new mixing matrix eliminating CMB, i.e.,
\begin{equation}
 \tilde{\bm D}\equiv
     \begin{pmatrix}
\bm{D}^{\rm dust}(\bar{p}^I)&\bm{D}^{\rm
    dust}_{,1}&\cdots&\bm{D}^{\rm dust}_{,N}&\bm{D}^{\rm synch}(\bar{p}^J)&\bm{D}^{\rm
    synch}_{,1}&\cdots&\bm{D}^{\rm synch}_{,M}&
    \cdots
     \end{pmatrix}\,.
\end{equation}
Note that $\tilde{\bm{D}}$ is {\it not} a square matrix and thus not
invertible. For example, when we 
have only dust and synchrotron as foregrounds, and dust and synchrotron
have $N$ and $M$ spectral parameters, respectively, $\tilde{\bm{D}}$
becomes a $2N_{\rm pix}N_{\rm freq}$-by-$2N_{\rm pix}(N+M+2)$ matrix;
however, to be able to solve for a cleaned CMB map we need $N_{\rm
freq}\ge N+M+3$. 

We maximize the marginalized likelihood part of
Eq.~(\ref{eq:tobemaximized}) (the first line on the right hand side)
with respect to the foreground sky signal 
to obtain the maximum likelihood value of $\vec{s}_{f}$ as
\begin{equation}
 \vec{s}_{f}^{\rm ML}
  =\left[\tilde{\bm{D}}^{\rm T}(\bm{S}^{\rm CMB}+\bm{N})^{-1}\tilde{\bm{D}}\right]^{-1}\tilde{\bm{D}}^{\rm
    T}(\bm{S}^{\rm CMB}+\bm{N})^{-1}\vec{m}\,.
\end{equation}
Note that $\vec{s}_{f}^{\rm ML}$ is {\it not} the maximum
likelihood value of Eq.~(\ref{eq:bayesianlikelihood}) but is the maximum
likelihood value of the CMB-marginalized likelihood,
Eq.~(\ref{eq:tobemaximized}), ignoring the prior term.

Inserting $\vec{s}_{f}^{\rm ML}$ back into
Eq.~(\ref{eq:tobemaximized}), we obtain the final expression for the
posterior distribution:
\begin{align}
 \nonumber
  -2\ln P(\bar p^I,\vec{s}^{\rm ML}_{f},\bm{S}&|\vec{m})
  =
\vec{m}^{\rm T}\left(\bm{S}^{\rm
		CMB}+\bm{N}\right)^{-1}\vec{m}\\
\nonumber
  &-
\left[\tilde{\bm{D}}^{\rm
    T}(\bm{S}^{\rm CMB}+\bm{N})^{-1}\vec{m}\right]^{\rm T}
\left[\tilde{\bm{D}}^{\rm T}(\bm{S}^{\rm CMB}+\bm{N})^{-1}\tilde{\bm{D}}\right]^{-1}\tilde{\bm{D}}^{\rm
T}(\bm{S}^{\rm CMB}+\bm{N})^{-1}\vec{m}\\
 &
  +\ln\left|2\pi(\bm{S}^{\rm
		CMB}+\bm{N})\right|
  -2\ln P(\vec{s}_{f}^{\rm ML},\bm{S}^{f})+\mbox{constant}\,.
  \label{eq:posterior}
\end{align}
This formula is the foundation of the Delta-map method. Once again, as
$\tilde{\bm{D}}$ no longer depends on sky directions $\hat n$, we can
evaluate the posterior in pixel, spherical harmonics, or wavelet space
by using $\vec{m}$, $\bm{S}^{\rm CMB}$, and $\bm{N}$ in the appropriate basis.

To understand this rather complicated expression, let us use an example
of a spatially uniform power-law component as foreground and work with
two maps at two frequencies. Then the first two lines yield
\begin{equation}
\mbox{CMB}^{\rm ML}\left[\bm{S}_0^{\rm CMB}+
\frac{g_{\nu_2}^2D^2_{\nu_2}{\bm N}_{\nu_1}+g_{\nu_1}^2D^2_{\nu_1}{\bm N}_{\nu_2}}{\left(g_{\nu_2}D_{\nu_2}-g_{\nu_1}D_{\nu_1}\right)^2}	    
\right]^{-1}\mbox{CMB}^{\rm ML}\,,
\end{equation}
where $\mbox{CMB}^{\rm ML}$ is given by Eq.~(\ref{eq:simplestML}). 
We find that this expression agrees with the likelihood we constructed
in Eq.~(\ref{eq:(cal L)}) with appropriate coefficients $\alpha_j$ for
this example, up to the diagonal noise (which we have not
added here yet). Therefore, we conclude that
Eq.~(\ref{eq:(cal L)}) gives the CMB-marginalized posterior distribution of
$\bar{p}^I$ and the cosmological parameters, with the
maximum likelihood estimate of foregrounds for the CMB-marginalized likelihood.

As the determinant term in Eq.~(\ref{eq:posterior}) does not depend on the
mixing matrix (hence $\bar{p}^I$), maximizing the posterior with
respect to $\bar{p}^I$ with a flat prior,
$P(\vec{s}_{f},\bm{S}^{f})=\mbox{constant}$, amounts to minimizing
$\chi^2$ as in Eq.~(\ref{eq:chi2min}). This justifies our algorithm
presented in Sect.~\ref{sec:implementation}. 

We could have marginalized Eq.~(\ref{eq:tobemaximized}) over
$\vec{s}_{f}$. Then it would also modify the determinant which, in
turn, would depend on $\bar p^I$. This would be the situation in which
we maximize Eq.~(\ref{eq:(cal L)}) including the determinant for $\bar
p^I$, $r$, and $s$ simultaneously. However, we find that this procedure
produces a biased result for $\bar p^I$. The same behavior was observed
in \cite{2009MNRAS.392..216S} for a flat prior on $\vec{s}_{f}$. An intuitive reason for this is that, when foregrounds are allowed to vary
to unreasonably large values by a flat prior, the marginalized
distribution has $\bar p^I$ that is far away from the maximum likelihood
values. Therefore, a better approach would be either to use
$\vec{s}^{\rm ML}_{f}$ instead of marginalizing over it (which we have
adopted here), or to marginalize over $\vec{s}_{f}$ with a reasonable
choice of $\bm{S}^{f}$ (see
\cite{2016A&A...588A.113V,vansyngelthesis:2014} for this approach), or
to marginalize over $\bm{S}^f$.

Note that the posterior distribution given in Eq.~(\ref{eq:posterior})
does not make distinction between ``CMB channels'', ``synchrotron
channels'', and ``dust channels''. This is a nice feature; while
a heuristic construction of cleaned CMB maps based on identifying
low- and high-frequency maps as foreground channels (discussed in
Sect.~\ref{sec:deltamap}) as well as its likelihood given in Eq.~(\ref{eq:(cal
L)}) is intuitive and easy to understand/interpret, it might
give readers an impression that there is some arbitrariness in choosing
which channels are CMB etc. Consider the case where we have far more
channels than the number of foreground components. Then, how should we
decide which ones are CMB and foreground channels? The posterior
distribution we have derived in this section tells us that this
arbitrariness does not exist: we can use Eq.~(\ref{eq:posterior}) to
combine all the data optimally to estimate $r$ and the mean foreground
parameters.

For the rest of this paper, we shall explore the simplest cases in which
the number of frequency channels is just enough to solve for one CMB map
and the necessary number of the mean foreground parameters. Then
Eq.~(\ref{eq:(cal L)}) is the appropriate likelihood. That is to say,
$N_{\rm freq}=N+M+3$. For more general cases
in which $N_{\rm freq}$ is greater, Eq.~(\ref{eq:posterior}) should
be used.

\section{Validation with realistic foreground maps}
\label{sec:validation}
\subsection{Sky model}
Our fiducial sky model consists of three components: polarized CMB,
synchrotron, and thermal dust emission. Later in Sect.~\ref{sec:ame}, we
also explore the effect of AME on cleaning synchrotron. 

For synchrotron, we use 
``SynchrtronPol-commander\_0256\_R2.00'' at $30$ GHz as a template,
which was derived from Planck maps using the Commander algorithm
\cite{2016A&A...594A..10P}. We extrapolate this map
to other frequencies assuming power-law emission with 
spatially-varying spectral indices $\beta_{\rm s}$.  The spatial
variation is taken from \cite{2008A&A...490.1093M}, which was derived
from WMAP 23~GHz and Haslam 408~MHz maps.

For dust component, we explore two possibilities: one-component and
two-component MBB dust. The latter is to test robustness of the Delta-map
method for more complex sky model; namely, we build the Delta-map method
assuming one component, and apply it to a sky map containing
two-component dust. Our starting point is the dust polarization template
``DustPol-commander\_1024\_R2.00'' at $353$ GHz, which was derived from
Planck maps using the Commander algorithm
\cite{2016A&A...594A..10P}. Extrapolation to other frequencies is done
by either one- or two-component MBB.

The one-component MBB is given by Eq.~(\ref{eq:MBBequation}), with 
maps of  $\beta_{\rm d}(\hat n)$ and $T_{\rm d}(\hat n)$ taken from
the Commander analysis given in \cite{2016A&A...594A..10P}.

A two-component MBB model is motivated by an observation that the
one-component model does not explain simultaneously the Rayleigh-Jeans
(microwave) and  Wien (infrared) regions of the dust spectrum
observed by the FIRAS and DIRBE instruments on board COBE, respectively
\cite{1999ApJ...524..867F}. A two-component MBB dust
model has been proposed to explain the observations. Physically, two
components might represent different dust grains (e.g., silicate- and
carbon-dominated grains) with different temperatures.
Our two-component model is based on \cite{2015ApJ...798...88M}. The
model assumes the following frequency dependence in brightness temperature:
\begin{equation}
  F_\nu= a_1 \left(\frac{\nu}{\nu_0}\right)^{\beta_1+1}
		     \frac{1}{e^{h\nu/k_BT_1(\hat n)}-1}
  +a_2 \left(\frac{\nu}{\nu_0}\right)^{\beta_2+1} \frac{1}{e^{h\nu/k_BT_2(\hat n)}-1}~,
  \label{eq:two_component_model}
\end{equation}
with $\nu_0 = c/\lambda_0$ and $\lambda_0=100~\mu{\rm m}$. This function
then gives $D_\nu=F_\nu/F_{\nu_*}$. Here, 
$a_1$, $a_2$, $\beta_1$, and $\beta_2$ are constant fitting
parameters, whereas $T_2(\hat n)$
(hot dust temperature) is a spatially-varying parameter. The remaining
parameter, $T_1(\hat n)$ (cold dust temperature), is derived from $T_2$ assuming
 thermal equilibrium with the same interstellar-radiation field
 \cite{1999ApJ...524..867F}, i.e.,
\begin{equation}
 [T_1(\hat{n})]^{4+\beta_1} = \frac{1}{q_r}\frac{\zeta(4+\beta_2)}{\zeta(4+\beta_1)}\frac{\Gamma(4+\beta_2)}{\Gamma(4+\beta_1)}\left(\frac{h\nu_0}{k_B}\right)^{\beta_1-\beta_2}[T_2(\hat{n})]^{4+\beta_2}~.
\label{eq:T_2}
\end{equation}
Here we have set the constant parameters as $\beta_2 = 2.82$, $\beta_1=1.63$,
$a_1=f_1q_r$, $a_2=(1-f_1)$ with
$f_1 = 0.0485$, and $q_r=8.219$ \cite{2015ApJ...798...88M}.
The spatially varying parameters $T_2(\hat n)$ and
$[Q^f,U^f]_{\nu_*}(\hat n)$ with $\nu_*=353$~GHz
are set according to the Planck results \cite{2015ApJ...798...88M}.

\subsection{Results}
We apply the Delta-map method to estimate $r$ from 1000 random
realizations of the CMB at a Healpix resolution of $N_{\rm
side}=4$. We do not include instrumental noise here but investigate the
effect of noise in Sect.~\ref{sec:noise}. We prepare two sets of
simulations: one with no tensor mode ($r_{\rm input}=0$) and another with
$r_{\rm input}=10^{-3}$.

We smooth foreground maps and resample them at low-resolution pixels of
$N_{\rm side}=4$ following the procedure described in
Sect.~\ref{sec:implementation}. We use the P06 mask defined by the WMAP
polarization analysis \cite{2007ApJS..170..335P} and implement it in the
likelihood evaluation following Appendix~\ref{sec:mask}.

We compare two combinations of frequencies:
$(40,60,140,230,280,340)$~GHz (Case A) and $(40,60,140,230,340,400)$~GHz
(Case B). We show the results for $\nu_{\rm
CMB}=140$~GHz in this paper but one can
choose other $\nu_{\rm CMB}$\footnote{As stressed at the end of in
Sect.~\ref{sec:posterior}, choosing a ``CMB channel'' makes sense only
when we have just enough frequency channels to solve for one CMB
map. For more general cases in which we have redundant frequencies,
Eq.~(\ref{eq:posterior}) should be used.}. We do not apply bandpass
average.
In this and following sections, we explore the likelihood in the
parameter ranges of $r=[0, 1]$, $\beta_{\rm s}=[-4.0, -2.0]$, $\beta_{\rm
d}=[1.0,2.0]$ and $T_{\rm d}=[5~{\rm K}, 40~{\rm K}]$ to find the 
best-fitting parameters for each realization. We find that all the
best-fitting parameters are well within these parameter ranges listed
above for all the 1000 realizations.

\begin{figure}[t]
 \centering\includegraphics[width=1.0\linewidth]{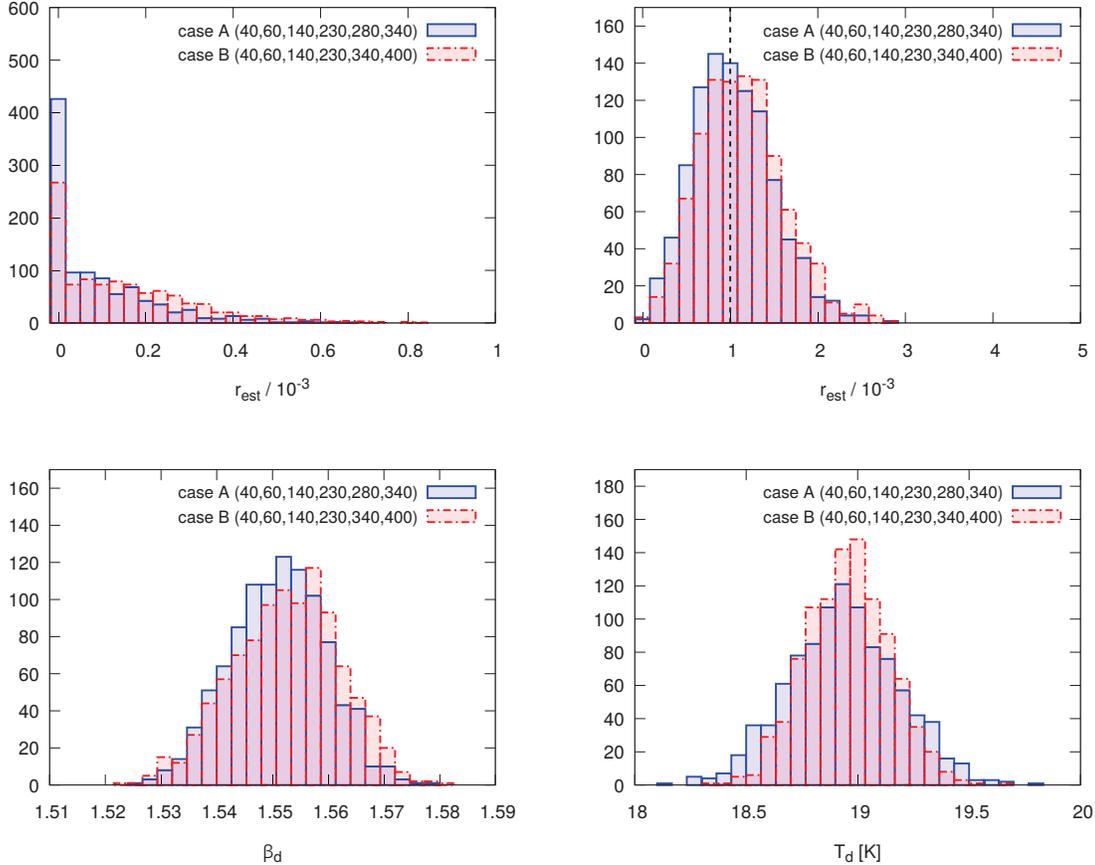}    
 \caption{Performance of the Delta-map method for power-law synchrotron
 and one-component MBB dust. (Upper panels) We show the distribution of
 $r$ estimated from 1000 random realizations of
 the CMB. No instrumental noise is included. The simulation also
 includes power-law synchrotron and one-component MBB dust.
  (Left) The input value of $r$ is zero. (Right) The input value of $r$
 is $10^{-3}$. We show the results for $\nu_{\rm CMB}=140$~GHz. The
 other channels are indicated in the figure. (Bottom panels) The
 distribution of $\bar\beta_{\rm d}$ and $\bar T_{\rm d}$ estimated from
 1000 random realizations, for the input value of $r=0$.} \label{fig:fig2}
\end{figure}

In Fig.~\ref{fig:fig2}, we show the results from applying the Delta-map
method for power-law synchrotron (Sect.~\ref{sec:2.2.1}) and
one-component MBB (Sect.~\ref{sec:MBB}) to CMB$+$power-law
synchrotron$+$one-component MBB simulations, i.e., the simulations and
assumptions match up to first order in spatial variation of the
foreground parameters.  For $r_{\rm input}=0$ (left panel), we find no
significant bias, with a slightly larger uncertainty in $r$ for Case
B. The uncertainties are dominated by cosmic variance of the
gravitational lensing effect \cite{2011ApJ...737...78K} and are found to
be $r<0.34 \times 10^{-3}$ for Case A and $r<0.48 \times 10^{-3}$ for
Case B (95\%~CL). These results are much better than our previous study
that found a significant bias for the P06 mask, $\delta r=2\times
10^{-3}$ \cite{2011ApJ...737...78K}, because our previous algorithm had
to assume that the spectral parameters are uniform across the sky. The
Delta-map method takes care of the spatial variation of the spectral
parameters, as promised.
Note that, for $r_{\rm input}=0$, a large number of samples fall
into the first bin because foregrounds are successfully removed and we
have put a hard prior of $r\geq 0$.

\begin{figure}[t]
 \centering\includegraphics[width=1.0\linewidth]{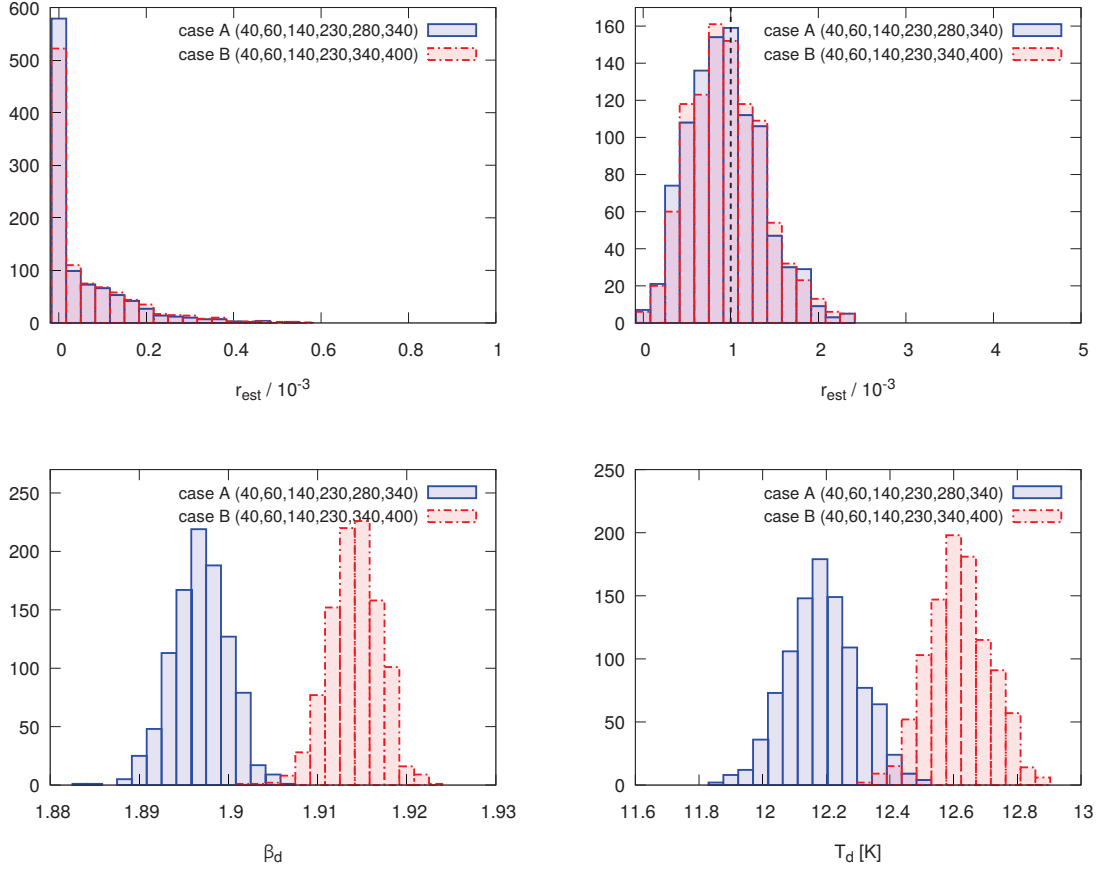}    
 \caption{Same as Fig.~\ref{fig:fig2}, but the Delta-map method is
 applied to simulations with CMB$+$power-law synchrotron$+$two-component
 MBB dust. Note that the Delta-map method still assumes that dust is
 one-component MBB. We show the results for $\nu_{\rm CMB}=140$~GHz. The
 other channels are indicated in the figure.}
 \label{fig:fig3}
\end{figure}

For $r_{\rm input}=10^{-3}$ (right panel), we find $r = (1.12 \pm 0.45)
\times 10^{-3}$ for Case A and $(1.22 \pm 0.48) \times 10^{-3}$ for Case
B (68\%~CL). The error bar here is dominated by both
CMB lensing and cosmic variance of the input tensor mode.

Next, we apply the Delta-map method for power-law synchrotron and
one-component MBB  to CMB$+$power-law synchrotron$+${\it two}-component
MBB simulations, i.e., the simulations and assumptions no longer match
for dust. Can we still recover the input $r$ without a significant bias?
In Fig.~\ref{fig:fig3} we show the recovered $r$. For $r_{\rm input}=0$
(left panel) we find $r<0.26\times 10^{-3}$ for Case A and $r<0.31\times
10^{-3}$ for Case B (95\%~CL). For $r_{\rm input}=10^{-3}$ (right panel)
we find $r=(1.02\pm 0.42)\times 10^{-3}$ for Case A and $(1.05\pm
0.43)\times 10^{-3}$ for Case B (68\%~CL). Interestingly we obtain
slightly a better result for this case, but the difference with respect
to the previous case is well within the error bar.

We thus conclude that the Delta-map method recovers the
input value of $r$ without a significant bias in the presence of
realistic spatially-varying foreground spectral parameters.
The slight difference in the uncertainties for Case A and B arises
from foreground residuals. Generally, in the absence of instrumental
noise we expect smaller uncertainties if the foreground channels are
closer to the CMB channel, because a mismatch of foreground
emission between the CMB and foreground channels decreases as the CMB
and foreground channels become closer to one another.

\subsection{Null test}
\label{sec:null}
To test our algorithm more precisely, we perform a null test;
namely, we clean two maps (100 and 140~GHz) following the procedure in
Sect.~\ref{sec:multiclean}, estimate $r$, and take the
difference to see if there is any residual bias in the difference. This is
a powerful null test (see \cite{2009ApJS..180..225H} for application
to the optical depth parameter $\tau$), as it is free from cosmic
variance, allowing for precise test with a smaller error bar. 
We must perform this kind of test on the real data to make sure that a
future detection of $r$ is not due to a residual foreground. 
As we do not include instrumental noise yet, the estimates of $r$ at
$100$~GHz ($r_{100}$) and $140$~GHz ($r_{140}$) should be identical if the
foreground emission is removed adequately.

\begin{figure}[t]
 \centering\includegraphics[width=1.0\linewidth]{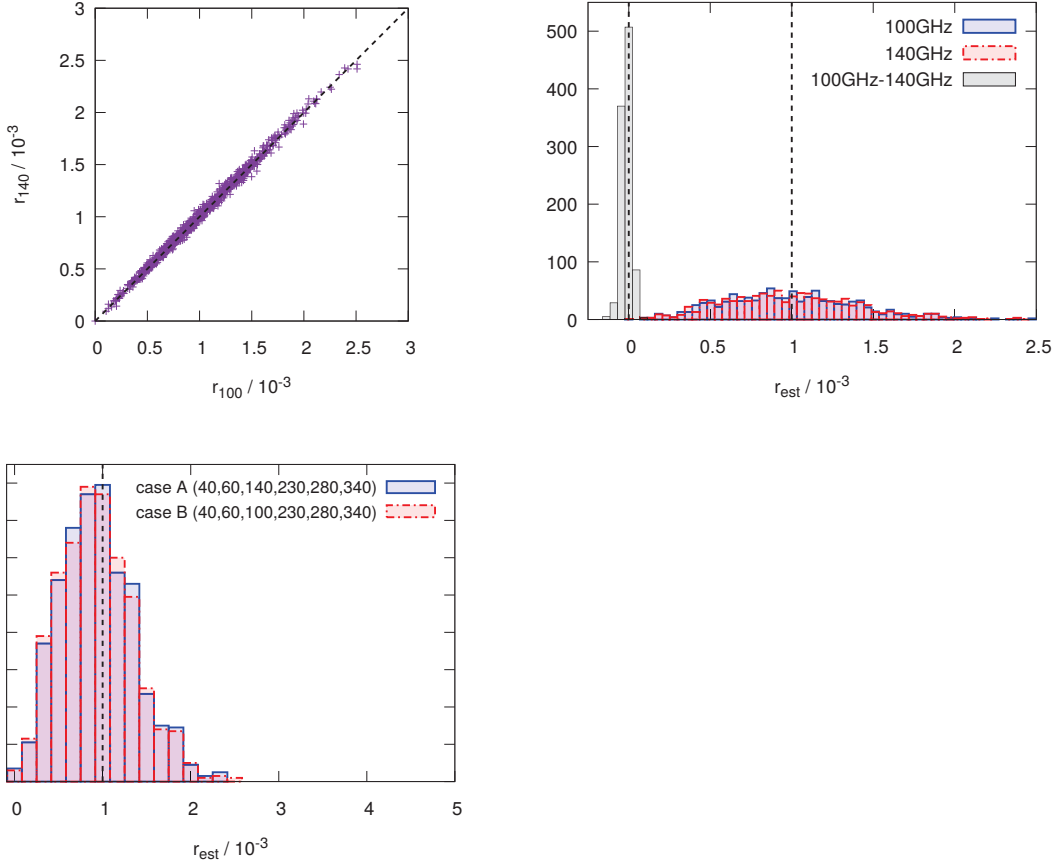}    
 \caption{Null test for $r$ measured from cleaned 100~GHz 
 ($r_{100}$) and 140~GHz ($r_{140}$) data. (Upper left) Scatter
 plot of $r_{100}$ and $r_{140}$. (Upper right) Distribution of $r_{100}$ (purple),
 $r_{140}$ (green), and $r_{100}-r_{140}$ (red). (Bottom left) A
 zoom-up of the distributions of $r_{100}$ and $r_{140}$ in the upper
 right panel, with the same
 bin size as in the top-right panel of Fig.~\ref{fig:fig3}. The input
 value of $r$ is $10^{-3}$.} \label{fig:fig4}
\end{figure}

We perform the null test for the simulations with $r_{\rm
input}=10^{-3}$, power-law synchrotron, and two-component MBB dust
(while the Delta-map method assumes one-component MBB). Frequency
channels are $(40,60,100,140,230,280,340)$~GHz.

In the left panel of Fig.~\ref{fig:fig4}, we show that $r_{100}$ and
$r_{140}$ are indeed highly correlated. In the right panel, we show the
distribution of the difference, $r_{100}-r_{140}$, as well as those of
$r_{100}$ and $r_{140}$. We find that the distribution is consistent
with the null hypothesis and that the small spread of the distribution of
$r_{100}-r_{140}$ is due to foreground residuals.
We find $r_{100}-r_{140}=(0.72\pm 2.98)\times 10^{-5}$ (68\%~CL).

\section{Applications}
\label{sec:application}
\subsection{Decorrelation}
Internal template methods assume that a map of $[Q,U]$ at one frequency
can be used to predict $[Q,U]$ at other frequencies as a function of
foreground parameters. However, this assumption breaks down and leads to
a phenomenon called ``decorrelation'', when we have a superposition of
multiple components with different foreground parameters; e.g.,
superposition of multiple dust components with different values of
$\beta_d$ and $T_d$ along the line of sight
\cite{2015MNRAS.451L..90T,2017PhRvD..95j3511P}. The same effect arises
when we pixelize (or coarse-grain) a map with big pixels containing many
smaller pixels with different values of $\beta_d$ and $T_d$. Currently
there is no evidence for the former, intrinsic decorrelation effect
\cite{2018arXiv180104945P}, whereas the latter effect can be modeled
given the distribution of foreground parameters on smaller pixels. As we
shall show in this section, one way to deal with this decorrelation
effect is to use the foreground parameters that are slightly
different for Stokes $Q$ and $U$.

\subsubsection{Theory}
Let us take dust as an example, though the results can be generalized
straightforwardly to other foreground components. Consider that we have
multiple dust components along the line of sight or within a big
pixel. Each $i$th component has a temperature $T_{{\rm d},i}$ and a
polarization direction $\gamma_i$. We then observe
\begin{eqnarray}
 Q_\nu&=&\sum_i Q_{\nu,i} =g_\nu\sum_i I_{{\rm d},\nu}(T_{{\rm d},i})\Pi_i\cos(2\gamma_i)\,,\\
 U_\nu&=&\sum_i U_{\nu,i} =g_\nu\sum_i I_{{\rm d},\nu}(T_{{\rm d},i})\Pi_i\sin(2\gamma_i)\,,
\end{eqnarray}
where $I_{{\rm d}}$ is an unpolarized dust amplitude in brightness
temperature units and $\Pi_i$ is a polarization fraction. We have omitted
the sky direction vector $\hat n$ to simplify notation. Here we assume
that it is only the dust temperature $T_{{\rm d},i}$ that varies and we
fix the dust emissivity $\beta_{{\rm d}}$, but it is straightforward to
include it in the final result.

We consider $Q$ first. Writing dust temperatures as $T_{{\rm d},i}=\bar{T}_{{\rm d}}\left(1+\delta_i\right)$, we obtain
\begin{equation}
 Q_\nu=g_\nu I_{{\rm d},\nu}(\bar{T}_{\rm d})\sum_i
  \left(1+\frac{x_{\rm d}e^{x_{\rm d}}\delta_i}{e^{x_{\rm d}}-1}+\cdots\right)\Pi_i\cos(2\gamma_i)\,, 
\end{equation}
where $x_{\rm d}\equiv h\nu/(k_B\bar T_{\rm d})$. Writing
$A\equiv \sum_i\Pi_i\cos(2\gamma_i)$, we obtain
\begin{equation}
 Q_\nu=g_\nu I_{{\rm d},\nu}(\bar{T}_{\rm d})A
  \left[1+\frac{x_{\rm d}e^{x_{\rm d}}}{e^{x_{\rm d}}-1}
\frac{\sum_i\delta_i\Pi_i\cos(2\gamma_i)}{\sum_i\Pi_i\cos(2\gamma_i)}+\cdots\right]\,.
\end{equation}
If we define a new dust temperature fluctuation variable
\begin{equation}
{\delta}_Q\equiv \frac{\sum_i\delta_i\Pi_i\cos(2\gamma_i)}{\sum_i\Pi_i\cos(2\gamma_i)}\,,
\end{equation}
and truncate expansion at first order in $\delta_i$, we find
$Q_\nu=g_\nu I_{{\rm d},\nu}({T}_Q)A$, where ${T}_Q\equiv \bar{T}_{\rm
d}(1+{\delta}_Q)$. Performing similar calculations for $U$, we find
$U_\nu=g_\nu I_{{\rm d},\nu}({T}_U)B$, where $B\equiv
\sum_i\Pi_i\sin(2\gamma_i)$, ${T}_U\equiv \bar{T}_{\rm d}(1+{\delta}_U)$, and
\begin{equation}
 {\delta}_U\equiv \frac{\sum_i\delta_i\Pi_i\sin(2\gamma_i)}{\sum_i\Pi_i\sin(2\gamma_i)}\,.
\end{equation}
Now, we can always write $A$ and $B$ as $A=\Pi\cos(2\gamma)$ and
$B=\Pi\sin(2\gamma)$ by defining $\Pi$ and $\gamma$ by
$\Pi\equiv\sqrt{A^2+B^2}$ and $\tan(2\gamma)\equiv B/A$.
We obtain the final result: 
\begin{eqnarray}
 Q_\nu&=&g_\nu I_{{\rm d},\nu}({T}_Q)\Pi\cos(2\gamma)\,,\\
 U_\nu&=&g_\nu I_{{\rm d},\nu}({T}_U)\Pi\sin(2\gamma)\,.
\end{eqnarray}
Therefore, to first order in temperature perturbation, $Q$ and $U$
behave as if they had slightly different dust temperatures.
This makes an observed polarization direction, $\gamma_{\rm obs}$,
different from $\gamma$ because
\begin{equation}
 2\gamma_{\rm obs}=\tan^{-1}\left[\frac{U(\nu)}{Q(\nu)}\right]
=\tan^{-1}\left[\frac{I_{{\rm d},\nu}(T_U)}{I_{{\rm d},\nu}(T_Q)}\tan(2\gamma)\right]\,.
\end{equation}
Importantly, {\it the observed polarization direction changes with
frequency}, i.e., $\gamma_{\rm obs}=\gamma_{\rm obs}(\nu)$. This is the
origin of frequency decorrelation.

We can account for variations in any other foreground parameters (e.g.,
$\beta_{\rm d}$, $\beta_{\rm s}$, $C_{\rm s}$) by using slightly
different values for $Q$ and $U$.

\subsubsection{Impact on $r$}
We include decorrelation in our two-component MBB dust
simulation as follows. We perturb temperature of hot dust, $T_2$, in
$Q$ and $U$ as
\begin{eqnarray}
 T^Q_2(\hat{n}) &\equiv& T_2(\hat{n}) + \epsilon_Q \sigma_T~,\\
 T^U_2(\hat{n}) &\equiv& T_2(\hat{n}) + \epsilon_U \sigma_T~, 
\label{eq:de-corr}
\end{eqnarray}
where $\epsilon_Q$ and $\epsilon_U$ are Gaussian random numbers with
unit variance and $\sigma_T$ is a parameter that controls the effect of
decorrelation.  Temperatures of cold dust, $T^Q_1$ and $T^U_1$, are
related to $T^Q_2$ and $T^U_2$ via Eq.~(\ref{eq:T_2}). Using these
temperatures, we extrapolate the Planck dust polarization maps in
brightness temperature units at 353~GHz to other frequencies using
\begin{eqnarray}
 Q_\nu(\hat{n}) &=& g_{\nu}\frac{I_{{\rm
  d},\nu}(T^Q_2(\hat{n}))}{I_{{\rm d},353}(T_2(\hat{n}))}Q^f_{353}(\hat{n})~,\\
 U_\nu(\hat{n}) &=& g_{\nu}\frac{I_{{\rm
  d},\nu}(T^U_2(\hat{n}))}{I_{{\rm d},353}(T_2(\hat{n}))}U^f_{353}(\hat{n})~,
\end{eqnarray}
where $I_\nu(T_2(\hat{n}))$ is the
frequency spectrum in the two component model given by
Eq.~(\ref{eq:two_component_model}).

We can implement $T^Q_2$ and $T^U_2$ directly into the Delta-map method
by increasing the number of mean temperatures from one $\bar{T}_{\rm d}$
to two $\bar{T}^Q_2$ and $\bar{T}^U_2$. However, in this section we instead
apply the Delta-map method for one temperature $\bar{T}_{\rm d}$ and
see how much impact the decorrelation has on our estimation of $r$. To
this end, we generate temperature fluctuations with r.m.s. of 1~K, i.e.,
$\sigma_T=1$~K and include them in the hot dust temperature of
two-component MBB dust. To avoid negative temperature by fluctuations,
we do not perturb it if $T_2< 5$~K. 

 \begin{figure}[t]
 \centering\includegraphics[width=1.0\linewidth]{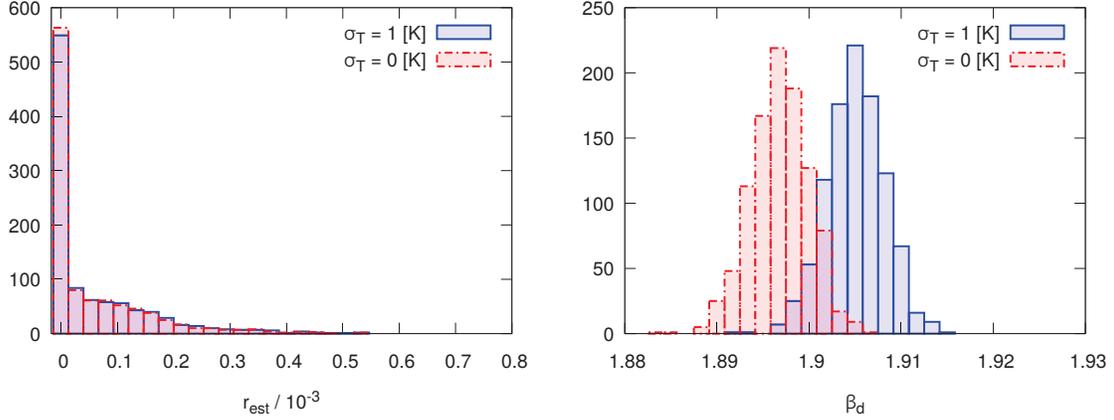}    
  \caption{Distribution of $r$ and $\bar{\beta}_{\rm d}$ obtained from
  1000 random realizations of the CMB with (purple) and without (green)
  decorrelation of dust polarization in the simulation. The Delta-map
  method assumes that there is no decorrelation.
  We used (40,60,230,280,340)~GHz channels to clean $140$~GHz
  channel. The simulation also includes power-law synchrotron and
  two-component MBB dust.}
  \label{fig:fig5}
 \end{figure}

In Fig.~\ref{fig:fig5} we find that decorrelation  with $\sigma_T=1$~K
shifts the best-fitting value of $\bar{\beta}_{\rm d}$ (right panel); a
superposition of varying foreground parameters changes the average
foreground spectrum as shown by \cite{2017MNRAS.472.1195C}. Fortunately,
this has little impact on $r$ (left panel). 
This is a good news for estimation of $r$, though the appropriate magnitude of
$\sigma_T$ in real sky and r.m.s. of other foreground parameters are yet to be
determined. At least our method provides a framework within which
frequency decorrelation due to any foreground parameters can be treated
consistently to first order in perturbation. In
table~\ref{table:foregroundparams}, we summarize the mean and variance of
the sky-averaged foreground parameters obtained from 1000 random
realizations of the CMB with the assumed foreground maps and the
frequency channels used in the analysis.

\begin{table}[t]
 \caption{Mean and variance of the reconstructed sky-averaged foreground parameters}
 \centering
 \begin{tabular}[t]{c|c|c|c}
  \hline\hline
 dust model ($r=0$)& $\bar\beta_s$ & $\bar\beta_d$ & $\bar T_d$ \\
 \hline
 one-component (case A) & $-2.920 \pm 0.020$ & $1.5523 \pm 0.009$ & $18.97 \pm 0.24$\\
 one-component (case B) & $-2.906 \pm 0.024$ & $1.5540 \pm 0.010$ & $18.99 \pm 0.19$\\
 two-component (case A) & $-2.940 \pm 0.026$ & $1.8977 \pm 0.003$ & $12.21 \pm 0.11$\\
 two-component (case B) & $-2.934 \pm 0.029$ & $1.9150 \pm 0.003$ & $12.64 \pm 0.10$\\
 two-component (case A with $\sigma_T=1$ [K]) & $-2.939 \pm 0.026$ &
	  $1.9059 \pm  0.003$ & $12.06 \pm 0.11$\\
	        \hline\hline
 \end{tabular}
   \label{table:foregroundparams}
\end{table}

\subsection{AME}
\label{sec:ame}
While it is widely believed that Galactic synchrotron and thermal dust
emissions are the most dominant sources of polarized foreground
emission on large angular scales \cite{2014PTEP.2014fB109I}, there
might be a third polarized component, AME, whose most plausible
candidate includes tiny PAH particles spinning with dipole moments
\cite{2013AdAst2013E..12L,2018NewAR..80....1D}. While the AME has been
shown to contribute to $\mu$K-level fluctuations in unpolarized
intensity at $40$~GHz and even smaller fluctuations at higher
frequencies \cite{2013ApJS..208...20B}, little is known about its
polarization properties.

The Planck collaboration has put an upper bound on the polarization
fraction of AME of $\Pi \lesssim 1.6\%$ in the Perseus region \cite{2016A&A...594A..25P}. The
QUIJOTE collaboration has placed a limit of $\Pi \lesssim 0.39 \%$ in
the molecular complexes W43 \cite{2017MNRAS.464.4107G}. The polarization
degree in more diffuse areas that are suitable for cosmological analyses
is unknown owing to the strong contamination from Galactic synchrotron
emission. Even though the polarization contribution of AME to the
microwave data near the foreground minimum, $\nu\approx 80-100$~GHz, is
expected to be negligible, it affects component separation by changing
interpretation of low frequency data from pure synchrotron to
synchrotron plus AME. In this sense, AME affects estimation of $r$
indirectly. In this section we show that this indirect effect can be
absorbed into the synchrotron curvature parameter $C_{\rm s}$.

To estimate the effect of AME, we model it following
\cite{2016MNRAS.458.2032R}. Specifically, we take the intensity map of
thermal dust emission at $353$~GHz and rescale it by a factor of 0.9 that is
found in the Planck 2015 results \cite{2016A&A...594A..25P} to predict
the AME intensity at 22.8~GHz. We then assign a 1\% polarization fraction
with the same polarization angles as those of the thermal dust
component, with a frequency spectrum given by the cold-medium model described in
\cite{2009MNRAS.395.1055A}.

We first apply the Delta-map method to the simulation with
CMB$+$power-law synchrotron$+$one-component MBB dust$+$AME,
using the Delta-map method for power-law synchrotron and one-component
MBB dust. We use $(40,60,140,230,280,340)$~GHz. We find a
significant bias in the estimated $r$: $r_{\rm est}=0.068 \pm 0.011$
(68\%~CL) for $r_{\rm input}=0$. At first this result was surprising because
the amplitude of AME at 140~GHz is too small to be relevant. We then
find that AME affects $r$ via changing the best-fitting synchrotron.
We therefore added curvature to the synchrotron spectrum and implement
it in the Delta-map method following the procedure in
Sect.~\ref{sec:synchcurv}. Including $C_{\rm s}$ requires one more
frequency channel for synchrotron, so we add one more frequency, which
is to be determined. When adding $C_{\rm s}$, we need to specify the
pivot frequency $\nu_{\rm s}$; we write a curved power-law synchrotron
spectrum as $\nu^{\beta_{\rm s}+C_{\rm s}\ln(\nu/\nu_{\rm s})}$.
We try $\nu_{\rm s}=40$ and $50$~GHz. Choice of $\nu_{\rm s}$ does not
affect the CMB result because it cancels in the cleaned CMB map, but it
would change the meaning of synchrotron parameters.

\begin{figure}[t]
    \centering\includegraphics[width=1.0\linewidth]{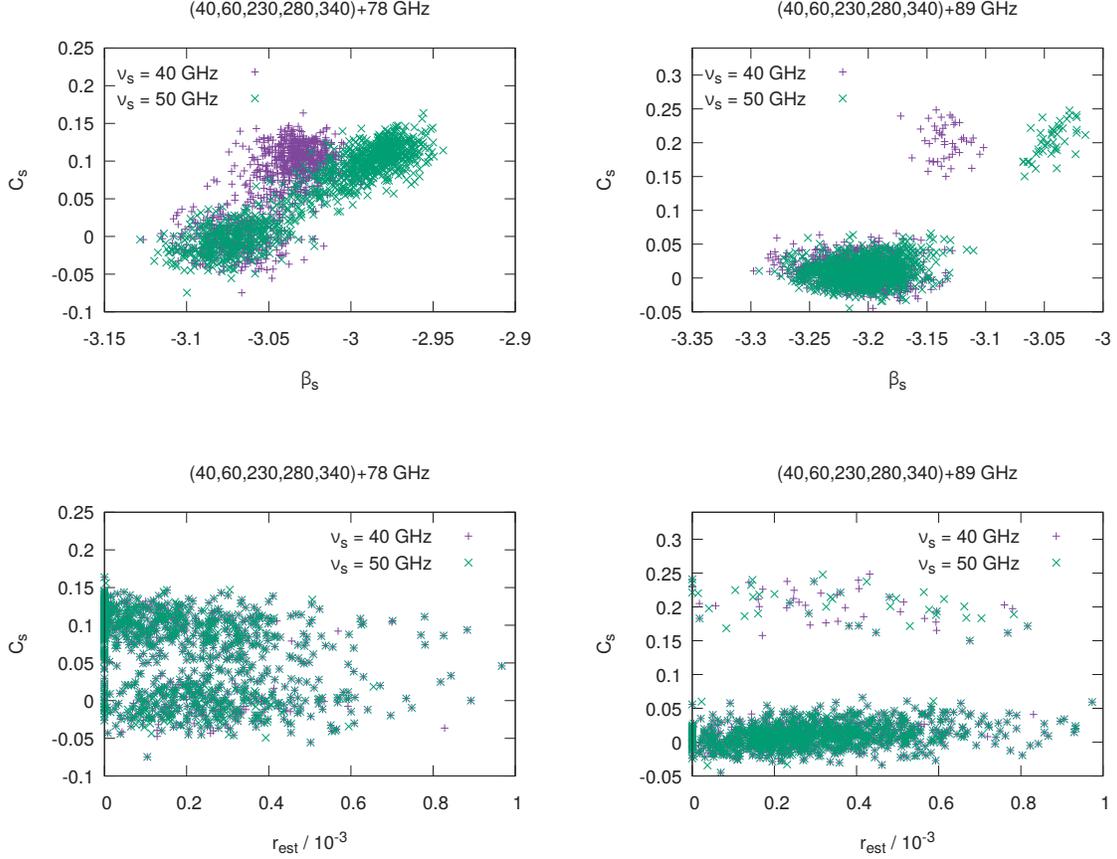}
 \caption{Scatter plots of $\bar{\beta}_{\rm s}$ and $\bar{C}_{\rm
 s}$ (top panels) and $r$ and $\bar{C}_{\rm s}$ (bottom panels) obtained
 from 1000 random realizations of the CMB with $r=0$. The simulation
 also includes AME, power-law synchrotron, and one-component MBB
 dust. We apply the Delta-map method for curved power-law synchrotron
 and one-component MBB dust. We show the results for $\nu_{\rm
 CMB}=140$~GHz. The other frequencies used are shown on top of each
 panel. We consider two pivot frequencies ($40$ and $50$ GHz), as shown
 in the figure.}
 \label{fig:fig6} 
\end{figure}
\begin{figure}[t]
  \centering\includegraphics[width=1.0\linewidth]{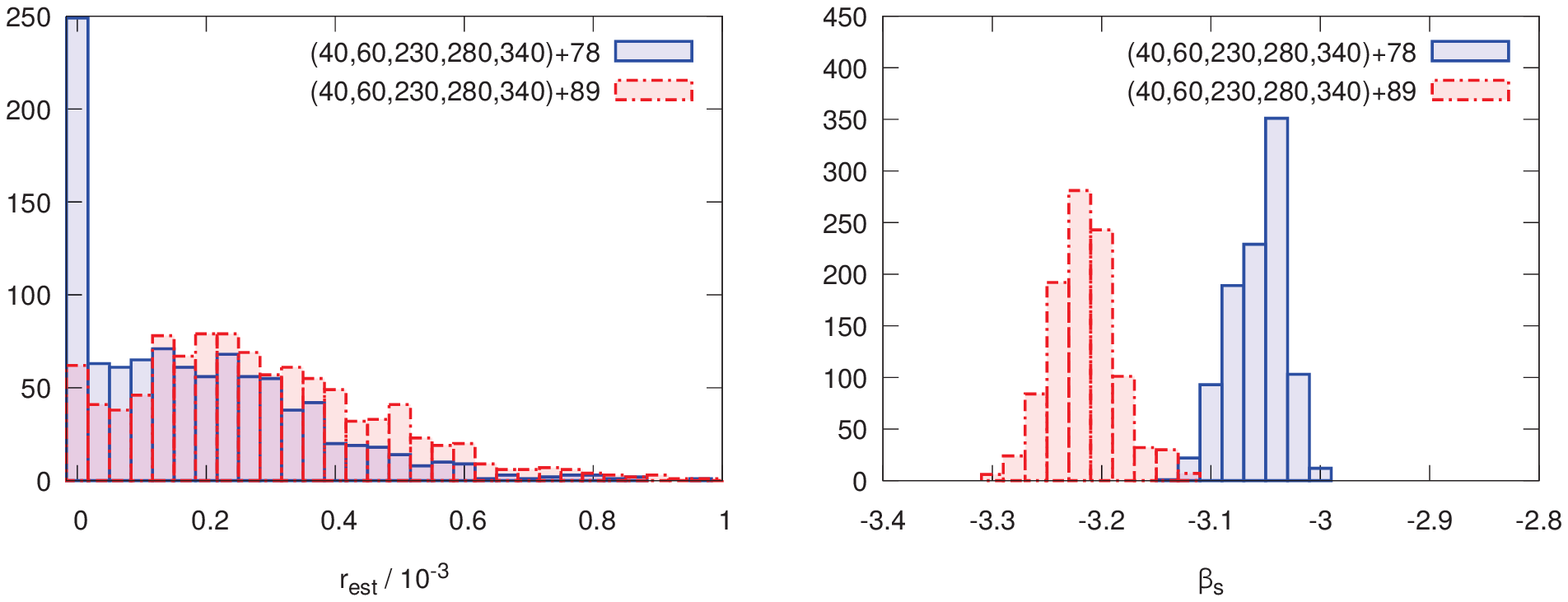}    
 \caption{Distribution of $r$ (left) and $\bar{\beta}_{\rm s}$
   (right) estimated from 1000 random realizations of the CMB with
 $r=0$. The simulation also includes AME, power-law synchrotron, and
 one-component MBB dust. We use the Delta-map method for one-component
 MBB dust and curved power-law synchrotron with the pivot frequency of
 $\nu_{\rm s}=40$ GHz.  We show the results for $\nu_{\rm CMB}=140$~GHz.}
   \label{fig:fig7}
\end{figure}

In the top panels of Fig.~\ref{fig:fig6}, we show scatter plots of $\bar{C}_{\rm
s}$ and $\bar{\beta}_{\rm s}$ from 1000 simulations of CMB$+$AME$+$power-law synchrotron$+$one-component MBB dust emission.
We study four configurations: to our Case A
 configuration ($40,60,140,230,280,340$)~GHz, we add 78 GHz (left panel)
 or 89 GHz (right panel) for $\bar{C}_{\rm s}$, and for the pivot scale
 we choose $\nu_{s}=40$~GHz or $50$~GHz. 
 We find that the distribution is split into two clusters. One clusters
 around $\bar{C}_{\rm s}\simeq 0$ and $\bar{\beta}_{\rm s}\simeq -3.07$
 and $-3.20$ in the left and right panels, respectively, while the other around
 $\bar{C}_{\rm s}=0.1$ and $0.2$ in the left and right panels, respectively.
 In the former cluster, because $|\bar{\beta}_{\rm s}|\gg |\bar{C}_{\rm
s}\ln(\nu/\nu_{\rm s})|$, most of the low-frequency foreground emission can be
removed as a power-law emission, and AME is absorbed by the Delta-map
terms proportional to $\delta\beta_{\rm s}(\hat{n})$ and
$\delta C_{\rm s}(\hat{n})$ (see Eq.~(\ref{eq:runningQU})). This
interpretation is supported by the fact that the distribution in this
cluster is not very sensitive to the choice of $\nu_{\rm s}$. In some
realizations, however, non-zero $\bar{C}_{\rm s}$ is needed by chance;
in this case the results fall into the latter cluster.

How about estimates of $r$? In the bottom panels of Fig.~\ref{fig:fig6},
we find that the mean curvature parameter $\bar{C}_{\rm s}$ is not
correlated with $r$. In the left panel of Fig.~\ref{fig:fig7}, we show
the distribution of $r$ for two frequency configurations with $r_{\rm
input}=0$.  We find $r < 5.2\times 10^{-4}$ and $r< 6.4\times 10^{-4}$
(95\%~CL) for additional $78$ and $89$ GHz channel, respectively; thus,
we find no evidence for bias unlike when we did not include $C_{\rm
s}(\hat n)$ to absorb the effect of AME. In fact, the different
number of realizations in the first bin in the left panel of
Fig.~\ref{fig:fig7} is due to foreground residuals. Because the
polarization intensity of the AME rapidly decreases above 40 GHz, an
additional channel at 78 GHz does the better job than 89 GHz in order to
capture the spectrum of the AME component, and relatively larger
foreground residuals remain for an additional channel at 89 GHz.

On the other hand, the distribution of $\bar{\beta}_{\rm s}$ shifts
significantly between the
cases of adding $78$ or $89$ GHz. The peak positions in the distribution
of $\bar{\beta}_{\rm s}$ depend upon the frequency channels used to
absorb AME and the frequency spectrum of AME.  Because of the fact that
the AME spectrum is generally characterized by a negative spectral index
and a negative running index, the Delta-map method returns smaller
$\bar{\beta}_{\rm s}$ if we choose a higher frequency channel to absorb
AME (farther away from the other two low-frequency channels at $40$ and
$60$ GHz).  Even though the results for $\bar{\beta}_{\rm s}$ are
different (indicating that the physical meaning of $\bar{\beta}_{\rm s}$
is no longer a simple synchrotron index), we find no bias in the
estimated $r$ in either case. (We find a slightly better result for an
additional channel at 78~GHz as described above.) This is a good news for estimation of
$r$.

\section{Effect of instrumental noise}
\label{sec:noise}
All the results we have presented so far were derived without
instrumental noise. Here, we shall include instrumental noise in the
simulation. As with other template-cleaning methods, the
Delta-map method also suffers from a boosted noise; because the method is
based on a linear combination of maps at different frequencies having the
CMB signal in common, we cannot avoid removing some portion of the
CMB when removing foreground components from the CMB channel.  A linear
combination of Gaussian random noises, on the other hand, will simply
result in a larger Gaussian random noise; thus, the net effect
is to boost the noise relative to the CMB signal in the map after the
Delta-map method is applied.

The effective noise level in a cleaned CMB map can be readily
estimated once the noise level at each observing frequency channel is
specified. When we use two channels for synchrotron and three channels
for dust, the noise covariance matrix for each cleaned CMB map
is given by the noise terms in Eq.~(\ref{eq:covariancematrix}) as
${\bm n}^{\nu_{\rm CMB},{\rm cleaned}}=\left({\bm N}_{\nu_{\rm
  CMB}}+\sum_{j=1}^{5}\alpha_j^2 {\bm N}_{\nu_j}\right)/\left(1+\sum_{j=1}^5\alpha_j\right)^2$,
where $\alpha_i$ are the coefficients given by solving
Eqs.~(\ref{eq:alphas1}) and (\ref{eq:alphas2}), and ${\bm N}_{\nu}$ is
the noise covariance matrix at a frequency channel $\nu$.

In Table~\ref{table:simple}, we show some examples of noise
specifications and derived constraints on $r$ using them. The effective
noise levels we find are $3.0$, $4.0$, $3.9$, $3.7$, $3.4$, $3.6$,
$3.8$, $3.9$, $4.7$, $4.5$, and $7.0$~$\mu{\rm K}~{\rm arcmin}$
from the top to bottom specifications in the table.
\footnote{Note that these values
are derived with a Gaussian smoothing of $2220$ arcmin (FWHM). To obtain
the equivalent non-smoothed white noise amplitudes in units of  $\mu{\rm
K}~{\rm arcmin}$, we need to multiply these numbers by a factor of $3.76$.}  To
derive these values, we used the Delta-map method with a power-law
synchrotron (Sect.~\ref{sec:2.2.1}) with $\bar{\beta}_{\rm s}=-3.0$ and
one-component MBB dust (Sect.~\ref{sec:MBB}) with $\bar{\beta}_{\rm d}=
1.65$ and $\bar{T}_{\rm d}=19.15$~K.
The total effective noise slightly depends upon the foreground
parameters. In fact, recent results from S-PASS have shown that the
spectral index of Galactic synchrotron emission is slightly
steeper than this value  ($\bar{\beta}_{\rm s}^{\rm S-PASS}=-3.22$)
\cite{2018arXiv180201145K}; in that case, the total effective noise for
the specification described in the top row in Table~\ref{table:simple}
would change from $3.0$ to $2.65$ $\mu{\rm K}~{\rm arcmin}$.

\begin{table}[t]
 \caption{Some example noise specifications at 40, 60, 100, 140, 230,
 280, 340, and 400~GHz, and derived upper limits on $r$ from cleaned
 CMB maps after the Delta-map method. They are in units of thermodynamic
 $\mu{\rm K}~{\rm arcmin}$. The bold faces show the lowest
 values.}
 \centering
 \begin{tabular}{c|c|c|c|c|c|c|c|c|c}
 \hline\hline
  &\multicolumn{2}{c|}{synchrotron}&\multicolumn{2}{c|}{CMB}&\multicolumn{4}{c|}{dust}
  & limits on $r$ \\
\hline
  Frequency [GHz] & 40   & 60  & 100 & 140 & 230 & 280 & 340 & 400 & ($95$\% CL) \\ 
 \hline
  & 20.0 & 6.0 & -   & {\bf 1.0} & {\bf 1.0} & -   & 5.0 & 18.0 &
  ${\bf <1.4 \times 10^{-3}}$\\
  & 20.0 & 6.0 & -   & 2.0 & 2.0 & -   & 5.0 & 18.0 & $<2.1\times 10^{-3}$ \\
  & {\bf 10.0} & 6.0 & -   & 2.0 & 2.0 & -   & 5.0 & 18.0 &  $<2.0\times 10^{-3}$ \\
  & 20.0 & {\bf 3.0} & -   & 2.0 & 2.0 & -   & 5.0 & 18.0 &  $<1.9\times 10^{-3}$  \\
  & 20.0 & 6.0 & -   & {\bf 1.0} & 2.0 & -   & 5.0 & 18.0 &  $<1.6\times 10^{-3}$  \\
Noise%
  & 20.0 & 6.0 & -   & 2.0 & {\bf 1.0} & -   & 5.0 & 18.0 &  $<1.9\times 10^{-3}$  \\
  $\left[\mu \mbox{K arcmin}\right]$%
  & 20.0 & 6.0 & -   & 2.0 & 2.0 & -   & {\bf 2.5} & 18.0 &  $<1.9\times 10^{-3}$ \\
  & 20.0 & 6.0 & -   & 2.0 & 2.0 & -   & 5.0 &  {\bf 9.0} &  $<2.0 \times 10^{-3}$  \\
  & 20.0 & 6.0 & 2.0 &  -  & 2.0 & -   & 5.0 & 18.0 & $<2.9\times 10^{-3}$ \\
  & 20.0 & 6.0 & -   & 2.0 & -   & 2.0 & 5.0 & 18.0 & $<2.8\times 10^{-3}$  \\
  & 20.0 & 6.0 & -   & 2.0 & 2.0 & 2.0 & 5.0 & -    & $<5.7\times 10^{-3}$ \\
  \hline\hline
 \end{tabular}
  \label{table:simple}
   \end{table}

If we combine several cleaned CMB maps (e.g., 100 and 140~GHz), the effective
noise will be reduced to some extent and its inverse noise covariance
matrix will be given by the inverse weighted sum of the noise matrices
of the cleaned CMB maps as
\begin{equation}
 \left({\bm n}^{\rm tot}\right)^{-1} = \sum_{\nu_{\rm CMB}} \left({\bm n}^{\nu_{\rm CMB},{\rm cleaned}}\right)^{-1}~.
\end{equation}
Note that the noise cannot be reduced by as much as the inverse-square
of the number of the maps ($n_{\rm map}^{-1/2}$), because the noises in
the cleaned maps are correlated with one another. This might give
readers an impression that the Delta-map method as presented in this
paper is unbiased, but is not necessarily minimum variance.
However, this is not the case; as we have discussed at the end of
Sect.~\ref{sec:posterior}, the posterior distribution given in
Eq.~(\ref{eq:posterior}) can be used to combine all maps optimally
to estimate $r$ and the mean foreground parameters. 

\section{Conclusions}
\label{sec:conclusions}
In this paper we have extended the internal template method by
accounting for spatially-varying foreground parameters such as
$\beta_{\rm s}$, $C_{\rm s}$, $\beta_{\rm d}$, and $T_{\rm d}$. This was
achieved by expanding frequency spectra of foregrounds to first
order in perturbations (spatial variations)
\cite{1999NewA....4..443B,2016ApJ...829..113S}.
We show that a foreground-cleaned CMB map obtained
by our method (``Delta-map method'') is the maximum likelihood solution (Eq.~(\ref{eq:exactML})),
and its posterior distribution for the CMB parameters (e.g.,
tensor-to-scalar ratio parameter $r$) and the sky-averaged, mean
foreground parameters 
(e.g., $\bar{\beta}_{\rm s}$, $\bar{C}_{\rm s}$, $\bar{\beta}_{\rm d}$,
and $\bar{T}_{\rm d}$) is obtained by marginalizing over CMB and using
the maximum likelihood solution to the foreground maps (Eq.~(\ref{eq:posterior})). We have found
that the Delta-map method removes a bias in $r$, which was found in the
previous study assuming uniform spectral indices
\cite{2011ApJ...737...78K}, successfully. We have checked this against
realistic sky simulations including polarized CMB, power-law
synchrotron, and one- or two-component MBB dust emission. 

While we have formulated and validated the Delta-map method in pixel
space in this paper, it is valid in any other space such as spherical
harmonics and wavelet space. 

Armed with this new tool, we have presented physically motivated ways to
address the frequency decorrelation due to averaging of different foreground
parameters along the line of sight (or over pixels) and the effect of
AME on foreground cleaning. We have shown that, to first order in
perturbation, the decorrelation effect can be accounted for by using
slightly different foreground parameters for Stokes $Q$ and $U$. As an
application we simulated a dust map with the r.m.s. temperature
fluctuation along the line of sight of 1~K. We show that the best-fitting
dust index $\bar{\beta}_{\rm d}$ is affected by this but the estimation
of $r$ is hardly affected. 

While AME has negligible contribution to the foreground minimum,
80--100~GHz, it affects estimation of $r$ via its effect on synchrotron
cleaning. If not accounted for, it can result in a bias in $r$ as large
as $\delta r\simeq 0.06$ (for 1\% polarization in AME). However,
we have found that the effect of AME on synchrotron cleaning can be
absorbed by the curvature parameter in synchrotron spectrum, $C_{\rm
s}$, within the Delta-map method. We find that the Delta-map
method can reduce the bias in $r$ to undetected level in our simulation.

The remaining question is how the performance of the Delta-map method,
especially the derived uncertainty in $r$, compares with those of other
foreground cleaning methods in the literature. We leave this for future work.
\section*{Acknowledgment}
We thank Joint Study Group of the LiteBIRD collaboration for useful
discussion and feedback on this project. We thank M. Remazeilles for
making their AME maps available to us. EK thanks B.~D. Wandelt for
useful discussion on the relationship between the template fitting
method and the Bayesian inference detailed in
\cite{2016A&A...588A.113V,vansyngelthesis:2014}, and the trimester
program on ``Analytics, Inference, and Computation in Cosmology'' at
Institut Henri Poincar\'e for hospitality and stimulating discussion on
statistical inferences. We thank K. Mizukami, K. Natsume, and
T. Yamashita for their contributions to the earlier stage of this
project, and Y. Minami for checking some results and comments on the
draft. We also thank M. Mirmelstein and P. Motloch for comments on the
draft. This work was supported in part by JSPS KAKENHI Grant Numbers
JP15H05896, JP18K03616, JP16H01543, and JP15H05890, and JSPS
Core-to-Core Program, A. Advanced Research Networks.
\bibliographystyle{ptephy}
\bibliography{references}

\begin{thebibliography}{10}

\bibitem{2016ARAA..54..227K}
M.~{Kamionkowski} and E.~D. {Kovetz}, \araa, {\bf 54}, 227--269 (September
  2016),  {{1510.06042}}.

\bibitem{2015PhRvL.114j1301B}
{BICEP2/Keck and Planck Collaborations}, \prl, {\bf 114}(10), 101301 (March
  2015),  {{1502.00612}}.

\bibitem{2008A&A...491..597L}
S.~M. {Leach} et~al., \aap, {\bf 491}, 597--615 (November 2008),
  {{0805.0269}}.

\bibitem{2018JCAP...04..023R}
M.~{Remazeilles} et~al., \jcap, {\bf 4}, 023 (April 2018),  {{1704.04501}}.

\bibitem{2018arXiv180706208P}
{Planck Collaboration IV}, ArXiv e-prints (July 2018),  {{1807.06208}}.

\bibitem{2007ApJS..170..335P}
L.~{Page} et~al., \apjs, {\bf 170}, 335--376 (June 2007),
  {{astro-ph/0603450}}.

\bibitem{2009MNRAS.397.1355E}
G.~{Efstathiou}, S.~{Gratton}, and F.~{Paci}, \mnras, {\bf 397}, 1355--1373
  (August 2009),  {{0902.4803}}.

\bibitem{2011ApJ...737...78K}
N.~{Katayama} and E.~{Komatsu}, \apj, {\bf 737}, 78 (August 2011),
  {{1101.5210}}.

\bibitem{2012MNRAS.420.2162F}
R.~{Fern{\'a}ndez-Cobos}, P.~{Vielva}, R.~B. {Barreiro}, and
  E.~{Mart{\'{\i}}nez-Gonz{\'a}lez}, \mnras, {\bf 420}, 2162--2169 (March
  2012),  {{arXiv:1106.2016}}.

\bibitem{2016MNRAS.459..441F}
R.~{Fern{\'a}ndez-Cobos}, A.~{Marcos-Caballero}, P.~{Vielva},
  E.~{Mart{\'{\i}}nez-Gonz{\'a}lez}, and R.~B. {Barreiro}, \mnras, {\bf 459},
  441--454 (June 2016),  {{1601.01515}}.

\bibitem{2014PTEP.2014fB109I}
K.~{Ichiki}, Progress of Theoretical and Experimental Physics, {\bf 2014}(6),
  06B109 (June 2014).

\bibitem{1999NewA....4..443B}
F.~R. {Bouchet} and R.~{Gispert}, \na, {\bf 4}, 443--479 (September 1999),
  {{astro-ph/9903176}}.

\bibitem{2016ApJ...829..113S}
R.~{Saha} and P.~K. {Aluri}, \apj, {\bf 829}, 113 (October 2016),
  {{1509.03656}}.

\bibitem{2016A&A...594A..13P}
{Planck Collaboration XIII}, \aap, {\bf 594}, A13 (September 2016),
  {{1502.01589}}.

\bibitem{2005ApJ...622..759G}
K.~M. {G{\'o}rski}, E.~{Hivon}, A.~J. {Banday}, B.~D. {Wandelt}, F.~K.
  {Hansen}, M.~{Reinecke}, and M.~{Bartelmann}, \apj, {\bf 622}, 759--771
  (April 2005),  {{astro-ph/0409513}}.

\bibitem{2016SPIE.9904E..0XI}
H.~{Ishino} et~al.,
\newblock {LiteBIRD: lite satellite for the study of B-mode polarization and
  inflation from cosmic microwave background radiation detection},
\newblock In {\em Space Telescopes and Instrumentation 2016: Optical, Infrared,
  and Millimeter Wave}, volume 9904 of {\em \procspie}, page 99040X (July
  2016).

\bibitem{2007ApJ...656..641E}
H.~K. {Eriksen}, G.~{Huey}, R.~{Saha}, F.~K. {Hansen}, J.~{Dick}, A.~J.
  {Banday}, K.~M. {G{\'o}rski}, P.~{Jain}, J.~B. {Jewell}, L.~{Knox}, D.~L.
  {Larson}, I.~J. {O'Dwyer}, T.~{Souradeep}, and B.~D. {Wandelt}, \apj, {\bf
  656}, 641--652 (February 2007),  {{astro-ph/0606088}}.

\bibitem{2009MNRAS.392..216S}
R.~{Stompor}, S.~{Leach}, F.~{Stivoli}, and C.~{Baccigalupi}, \mnras, {\bf
  392}, 216--232 (January 2009),  {{0804.2645}}.

\bibitem{2016A&A...588A.113V}
F.~{Vansyngel}, B.~D. {Wandelt}, J.-F. {Cardoso}, and K.~{Benabed}, \aap, {\bf
  588}, A113 (April 2016),  {{1409.0858}}.

\bibitem{vansyngelthesis:2014}
F.~{Vansyngel},
\newblock {\em The blind Bayesian approach to Cosmic Microwave Background data
  analysis},
\newblock PhD thesis, Universit\'e Pierre et Marie Curie - Paris VI (December
  2014).

\bibitem{2016A&A...594A..10P}
{Planck Collaboration X}, \aap, {\bf 594}, A10 (September 2016),
  {{1502.01588}}.

\bibitem{2008A&A...490.1093M}
M.-A. {Miville-Desch{\^e}nes}, N.~{Ysard}, A.~{Lavabre}, N.~{Ponthieu}, J.~F.
  {Mac{\'{\i}}as-P{\'e}rez}, J.~{Aumont}, and J.~P. {Bernard}, \aap, {\bf 490},
  1093--1102 (November 2008),  {{0802.3345}}.

\bibitem{1999ApJ...524..867F}
D.~P. {Finkbeiner}, M.~{Davis}, and D.~J. {Schlegel}, \apj, {\bf 524}, 867--886
  (October 1999),  {{astro-ph/9905128}}.

\bibitem{2015ApJ...798...88M}
A.~M. {Meisner} and D.~P. {Finkbeiner}, \apj, {\bf 798}, 88 (January 2015),
  {{1410.7523}}.

\bibitem{2009ApJS..180..225H}
G.~{Hinshaw} et~al., \apjs, {\bf 180}, 225--245 (February 2009),
  {{arXiv:0803.0732}}.

\bibitem{2015MNRAS.451L..90T}
K.~{Tassis} and V.~{Pavlidou}, \mnras, {\bf 451}, L90--L94 (July 2015),
  {{arXiv:1410.8136}}.

\bibitem{2017PhRvD..95j3511P}
J.~{Poh} and S.~{Dodelson}, \prd, {\bf 95}, 103511 (May 2017).

\bibitem{2018arXiv180104945P}
{Planck Collaboration XI}, ArXiv e-prints, page arXiv:1801.04945 (January
  2018),  {{arXiv:1801.04945}}.

\bibitem{2017MNRAS.472.1195C}
J.~{Chluba}, J.~C. {Hill}, and M.~H. {Abitbol}, \mnras, {\bf 472}, 1195--1213
  (November 2017).

\bibitem{2013AdAst2013E..12L}
E.~M. {Leitch} and A.~C.~R. {Readhead}, Advances in Astronomy, {\bf 2013},
  352407 (2013).

\bibitem{2018NewAR..80....1D}
C.~{Dickinson} et~al., New Astronomy Reviews, {\bf 80}, 1--28 (February 2018).

\bibitem{2013ApJS..208...20B}
C.~L. {Bennett} et~al., \apjs, {\bf 208}, 20 (October 2013),  {{1212.5225}}.

\bibitem{2016A&A...594A..25P}
{Planck Collaboration XXV}, \aap, {\bf 594}, A25 (September 2016),
  {{1506.06660}}.

\bibitem{2017MNRAS.464.4107G}
R.~{G{\'e}nova-Santos}, J.~A. {Rubi{\~n}o-Mart{\'{\i}}n},
  A.~{Pel{\'a}ez-Santos}, F.~{Poidevin}, R.~{Rebolo}, R.~{Vignaga}, E.~{Artal},
  S.~{Harper}, R.~{Hoyland}, A.~{Lasenby}, E.~{Mart{\'{\i}}nez-Gonz{\'a}lez},
  L.~{Piccirillo}, D.~{Tramonte}, and R.~A. {Watson}, \mnras, {\bf 464},
  4107--4132 (February 2017),  {{1605.04741}}.

\bibitem{2016MNRAS.458.2032R}
M.~{Remazeilles}, C.~{Dickinson}, H.~K.~K. {Eriksen}, and I.~K. {Wehus},
  \mnras, {\bf 458}, 2032--2050 (May 2016),  {{1509.04714}}.

\bibitem{2009MNRAS.395.1055A}
Y.~{Ali-Ha{\"i}moud}, C.~M. {Hirata}, and C.~{Dickinson}, \mnras, {\bf 395},
  1055--1078 (May 2009),  {{0812.2904}}.

\bibitem{2018arXiv180201145K}
N.~{Krachmalnicoff}, E.~{Carretti}, C.~{Baccigalupi}, G.~{Bernardi},
  S.~{Brown}, B.~M. {Gaensler}, M.~{Haverkorn}, M.~{Kesteven}, F.~{Perrotta},
  S.~{Poppi}, and L.~{Staveley-Smith}, ArXiv e-prints (February 2018),
  {{1802.01145}}.

\end{thebibliography}
\appendix
\section{Mask}
\label{sec:mask}
One advantage of pixel-based internal-template methods is that one can
easily and exactly account for the effect of a mask through the noise
covariance matrix \cite{2007ApJS..170..335P}. To implement it, we
rewrite the likelihood function as
\begin{equation}
 {\cal L}=\frac{\exp[-\frac{1}{2}(\bm{N}^{-1}\vec{x})^{\rm T}
  (\bm{N}^{-1}\bm{S}\bm{N}^{-1}+\bm{N}^{-1})^{-1}(\bm{N}^{-1}\vec{x})]|\bm{N}^{-1}|}{|\bm{N}^{-1}\bm{S}\bm{N}^{-1}+\bm{N}^{-1}|^{1/2}}~,
\end{equation}
where $\bm{S}={\bm C}^{\rm tens}(r)+{\bm C}^{\rm scal}(s)$ and $\bm{N}$
are CMB signal and noise covariance matrices, respectively.

We rearrange the data vector and the covariance matrix such that
unmasked pixels appear first and masked pixels afterward.
Suppose that the structure of $\bm{N}$ is given by
\begin{equation}
 \bm{N}= 
\begin{pmatrix}
 T & U\\
 U^{\rm T} & W\\
\end{pmatrix},
\end{equation}
where $T$ is the noise covariance matrix for unmasked pixels,
$W$ is for masked pixels, and $U$ is for their
correlation. Then the matrix inverse is given by
\begin{equation}
 \bm{N}^{-1}= 
\begin{pmatrix}
 T & U\\
 U^{\rm T} & W\\
\end{pmatrix}
^{-1}=
\begin{pmatrix}
 T^{-1}+T^{-1}UQ^{-1}U^{\rm T}T^{-1} & -T^{-1}UQ^{-1} \\
 -Q^{-1}U^{\rm T}T^{-1} & Q^{-1} \\
\end{pmatrix}
\equiv
\begin{pmatrix}
 B_{11} & B_{12} \\
 B_{21} & B_{22} \\
\end{pmatrix},
\end{equation}
where $Q\equiv W-U^{\rm T}T^{-1}U$ and $T^{-1}=B_{11}-B_{12}B^{-1}_{22} B_{21}$
(Schur complement). We then assign infinite noise to the masked pixels such that $N\to N+\lambda(I-M)$, where $I$ is the unit
matrix and $M$ is the diagonal matrix whose elements are zero for masked
pixels and unity otherwise. In the limit of $\lambda\to \infty$, $Q \to
\infty$, and the inverse of $\bm{N}$ is given by
\begin{equation}
 \bm{N}^{-1} \to
\begin{pmatrix}
    T^{-1} && 0 \\
         0 && 0 \\
\end{pmatrix}
   =
\begin{pmatrix}
    B_{11} - B_{12}B_{22}^{-1}B_{21} && 0 \\
         0 && 0 \\
\end{pmatrix}.
\end{equation}

\begin{figure}[t]
    \centering\includegraphics[width=1.0\linewidth]{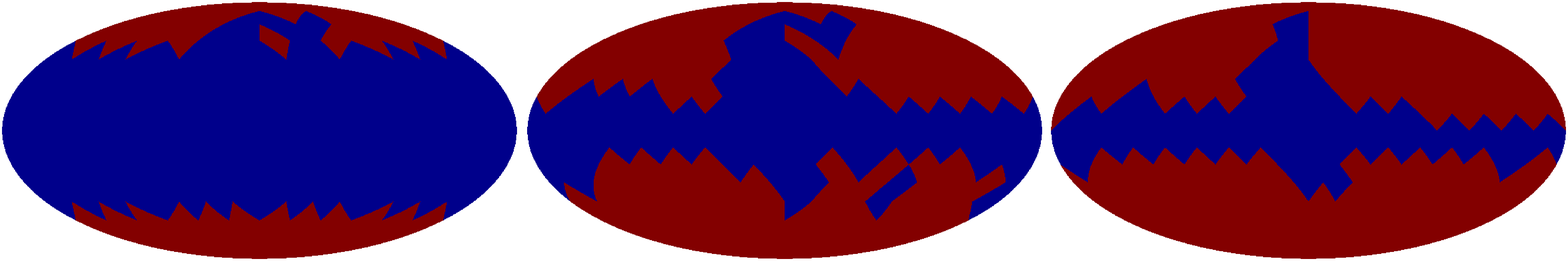}
   \caption{Conservative (left), fiducial (middle), and aggressive
 (right) masks used in this work, whose sky coverages  are $f_{\rm fsky}=0.22$, $0.59$, and $0.69$, respectively.}
     \label{fig:appendix}
\end{figure}

In this work, we have utilized the ``P06 mask'' defined by the WMAP polarization
analysis, degrading it to $N_{\rm side}=4$. The sky coverage of our
default P06 mask at $N_{\rm side}=4$ is $f_{\rm sky}=0.56$. Whilst we
find that this mask 
is sufficient to suppress the residual ${\cal O}(\delta p^2)$ term in
Eq.~(\ref{eq:QUmodel2}), it is interesting to investigate how the
results change if one uses different masks.  We apply the Delta-map
method to the simulations with two different masks, namely more
aggressive and conservative ones, whose sky coverages are $f_{\rm sky}=0.22$ and
$0.69$, respectively, as shown in Fig.~\ref{fig:appendix}. 
The results are summarized in Table~\ref{table:results_mask}. We find
that the aggressive mask leads to a larger bias and a weaker constraint
because of foreground residuals, while the conservative mask leads to a
smaller bias and a weaker constraint because of the smaller sky
coverage. We caution, however, that one should not accept the results for the
aggressive (and full-sky) case at face value because our foreground
model may not be accurate enough close to the Galactic plane.

\begin{table}[t]
 \caption{Limits on $r$ with different masks against the foreground
 model of one-component
 MBB dust and power-law synchrotron emissions with our case~A configuration
 ($40,60,140,230,280,340$ GHz). The input tensor-to-scalar ratio is $r=0$.}
 \centering
 \begin{tabular}[t]{c|c|c|c}
  \hline\hline
 mask & $f_{\rm sky}$ & mean & limit (95\% CL) \\
 \hline
 full-sky & 1.0 & $3.4 \times 10^{-4}$ & $ < 0.72 \times 10^{-3}$ \\
 aggressive & 0.69 & $1.7\times 10^{-4}$ & $< 0.48 \times 10^{-3}$\\
 fiducial & 0.59  & $9.9 \times 10^{-5}$ & $< 0.34 \times 10^{-3}$ \\
  conservative &0.22& $7.7 \times 10^{-5}$&$<0.52\times 10^{-3}$\\
  \hline\hline
 \end{tabular}
   \label{table:results_mask}
\end{table}
\end{document}